\documentclass[10pt,english,final,hidelinks,oneeqnum,onefignum,onetabnum,onethmnum]{siamart251216}
\usepackage[T1]{fontenc}
\usepackage[utf8]{inputenc}
\usepackage{units}
\usepackage{url}
\usepackage{amstext}
\usepackage{amssymb}

\makeatletter

\title{High-Performance Algorithms for Computing the V-BLAST Permutation\thanks{Funding: This research was supported in part by a grant from the Blavatnik
Computer Science Research Fund, and by the Israel Science Foundation
(grant 1524/23).}}

\author{Tomer Pelleg\thanks{Blavatnik School of Computer Science and AI, Tel Aviv University}\and{}Haim
Avron\thanks{School of Mathematical Sciences, Tel Aviv University}\and{}Sivan
Toledo\thanks{Blavatnik School of Computer Science and AI, Tel Aviv University\protect \\
Currently visiting ETH Zürich}}

\theoremstyle{nonumberplain}
\theoremheaderfont{\normalfont\scshape}
\theorembodyfont{\normalfont\itshape}
\theoremseparator{.}
\theoremsymbol{}
\newtheorem{lemmastar}{Lemma}
\theoremstyle{plain}

\headers{Computing the V-BLAST Permutation}
        {T. Pelleg, H. Avron, and S. Toledo}
\usepackage{placeins}
\makeatother

\usepackage{babel}
\begin{document}
\global\long\def\vblast{\text{V-BLAST}}%

\global\long\def\Span#1{\text{ Span}\left(#1\right)}%

\maketitle
\begin{abstract}
We propose two novel and efficient algorithms for computing a column
permutation called V-BLAST or an approximation of it. The V-BLAST
column permutation is used in a QR factorization that is part of an
integer/discrete least-squares solver. The overall solver is part
of the receiver in advanced communication systems that transmit multiple
symbols simultaneously. Two classes of efficient algorithms for computing
the exact V-BLAST permutation have been proposed in the literature.
The first, based on a sequence of Givens rotations, is unfortunately
highly sequential. The other, based on a sequence of downdating operations
on a pseudo-inverse, has a shorter critical path, but is unlikely
to be numerically stable. The new algorithms that we propose are inspired
by the similarity between V-BLAST and rank-revealing QR factorizations.
Our first algorithm uses Householder reflections, is numerically stable,
and has a short critical path. It is unable, however, to efficiently
exploit cache memories. We rectify this with a randomized algorithm
that computes an approximation to the V-BLAST permutation and has
a short critical path and can effectively exploit cache memories.
We prove approximation properties of the randomized algorithm and
demonstrate experimentally that both algorithms are faster than existing
ones, both when tested in isolation and when tested within an integer
least-squares solver.
\end{abstract}

\section{Introduction}

Several important applications, including wireless communication,
GPS positioning, and cryptanalysis, require the solution of integer
least-squares problems of the form
\[
\hat{x}=\arg\min_{x}\left\Vert W\left(Ax-b\right)\right\Vert _{2}
\]
where $A$ is a real or complex $m$-by-$n$ matrix with $m\geq n$,
$b$ is a real or complex vector, $W$ is a real matrix that represents
the weights of the $m$ constraints, and $x$ is either an integer
vector or a vector whose elements belong to a discrete set of real
or complex values~\cite{doi:10.1109/tit.2003.817444,doi:10.1109/tcomm.2003.809789,doi:10.48550/arxiv.2505.12980}.
In this paper, we assume that $A$ and $b$ have been pre-multiplied
by $W$, which allows us to assume from now on that $\hat{x}=\arg\min_{x}\|Ax-b\|_{2}$,
and also assume that $A$ has full column rank.

In communication, these problems are sometimes solved approximately
and sometimes exactly. In other domains, an exact minimizer is usually
required, or even the $k$ best solutions. Approximate solutions usually
return a so-called Babai point~\cite{doi:10.1007/bf02579403}, which
can be obtained by computing a QR factorization $A=QR$, where $Q$
is $m$-by-$n$ with orthonormal columns and $R$ is upper triangular,
followed by a substitution-like algorithm. The algorithm solves $R_{n,n}x_{n}=(Q^{*}b)_{n}$
and then rounds $x_{n}$ to the nearest admissible value $\tilde{x}_{n}$
(e.g., rounding to an integer; in communication, this is called quantization
or slicing). Now the problem is deflated by subtracting $R_{1:n-1,n}\tilde{x}_{n}$
from $(Q^{*}b)_{1:n-1}$. The process is then repeated until all the
elements of the approximate minimizer $\tilde{x}$ are resolved\footnote{We use Matlab notation for ranges of indexes, $i:j=\{i,i+1,\cdots,j\},$
where both $i$ and $j$ are integers; if $j<i$, then $i:j=\emptyset$.}.

To obtain a minimizer or the best $k$ solutions, a branch and bound
algorithm called the Schnorr-Euchner (SE) search is used~\cite{doi:10.1007/bf01581144}.
The SE search also requires the QR factorization of $A$. Most of
the details of the SE search are irrelevant in this article.

Both the approximation quality of the Babai point and the efficiency
of the SE search are strongly influenced by the particular matrix
$A$, whose columns are viewed in this context as the basis for a
lattice. Suppose that we transform the columns of $A$, 
\[
\hat{x}=\arg\min_{x}\left\Vert APP^{-1}x-b\right\Vert _{2}\;.
\]
If we choose a transformation $P$ that the set that $x$ belongs
to is invariant under, then we can solve the problem approximately
or exactly for $P^{-1}x$, searching in the domain of $x$, and then
recover a solution by multiplying by $P$.

When $x$ is an integer vector, $P$ can be any unimodular transformation.
An algorithm called LLL~\cite{doi:10.1007/bf01457454} or its variants
is often used in both GPS positioning and cryptanalysis to find a
unimodular matrix $P$ that is effective both in terms of the Babai
point and the efficiency of the SE search~\cite{doi:10.1109/tit.2013.2253596}.
However, in communication the elements of $x$ usually come from a
small finite set. These sets are, in general, not invariant under
unimodular transformations. In such cases, $P$ is usually restricted
to permutations~\cite{doi:10.1109/tit.2016.2627082}.

The dominant method for constructing a permutation $P$ that improves
the Babai point and the SE search is called V-BLAST~\cite{doi:10.1109/tit.2003.817444,doi:10.1109/issse.1998.738086}.
The original algorithms for constructing the V-BLAST permutation performed
$\Theta(mn^{2}+n^{4})$ arithmetic operations, but this was later
reduced to $\Theta(mn^{2}+n^{3})$. 

The next section describes both the original algorithm proposed for
constructing the V-BLAST permutation and a number of variants, some
much more efficient in terms of asymptotic operation counts. But on
modern hardware, operation counts often do not correlate well with
reduced running times. The ability to exploit multiple processing
units concurrently and the load on the memory system are often just
as significant, and often more significant.

Interestingly, to the best of our knowledge no effort has been made
to analyze the level of parallelism available in V-BLAST algorithms,
the load they present to the memory system (in terms of cache misses
and/or interprocessor communication) or to improve these metrics.

This paper addresses exactly these issues. We analyze parallelism
and cache misses in existing V-BLAST algorithms and show that most
of them are much worse in these aspects than most other cubic dense-matrix
algorithms. One of the existing V-BLAST algorithms offers reasonable
parallelism, but at the expense of reduced numerical stability.

We also propose two new algorithms: an exact V-BLAST algorithm with
good parallelism (similar to that in most cubic dense-matrix algorithms)
and excellent numerical stability, and an approximate V-BLAST algorithm
that offers both good parallelism and effective reuse of data in caches.
We analyze both algorithms theoretically and experimentally and show
that they are effective as claimed.

The rest of this article is organized as follows. The next section
provides additional background on V-BLAST and on existing algorithms
to compute it. Section~\ref{sec:Complexity-Analysis-of-Existing-Algorithms}
analyzes parallelism and cache misses in existing algorithms. Section~\ref{sec:VBLAST-via-Householder}
proposes a new algorithm to compute the V-BLAST permutation using
Householder reflections. Section~\ref{sec:R-V-BLAST} proposes a
randomized algorithm, based on matrix sketching, to compute an approximation
of the V-BLAST ordering. Section~\ref{sec:theoretical} establishes
certain important theoretical properties of the randomized algorithm,
effectively establishing that it produces a good approximation. Section~\ref{sec:Experimental-Results}
presents extensive experimental results that validate our theoretical
analyses, and Section~\ref{sec:Conclusions} presents our conclusions
from this research.

\section{Background}

In general, the aim of V-BLAST (and of a simpler method to construct
a permutation $P$ and of LLL and its variants) is to permute the
columns of $A$ such that its $R$ factor has large diagonal values
at the bottom right corner and small ones at the top left corner.
A particularly simple greedy method called SQRD is to compute a right-looking
$QR$ factorization with column pivoting using a Gram-Schmidt process.
In step $j$, the column with the smallest norm in the reduced matrix
is exchanged with column $j$, and its projection is subtracted from
columns to the right~\cite{doi:10.1049/el:20010899}. The magnitude
of $r_{jj}$, which is equal to the norm of the selected column in
the reduced matrix, is therefore minimized (greedily) in every step.
The column selection criterion is the exact opposite of the criterion
used in rank-revealing QR factorizations. This algorithm performs
$\Theta(mn^{2})$ operations. However, even if we ignore the numerical
instability of the Gram-Schmidt process, the permutation that this
method selects has been found to be inferior to the V-BLAST permutation~\cite{doi:10.1049/el:20010899,doi:10.1179/1752270611y.0000000004}.

V-BLAST is more sophisticated. It selects the columns from right to
left. At step $j$, it computes the norm of the orthogonal complement
of each remaining column with respect to all other remaining columns.
It selects the column with the largest orthogonal complement and permutes
it to position $n-j+1$. More formally~\cite{doi:10.1016/j.apnum.2007.01.008},
denoting the set of $i$ columns that appear last in the permutation
$P$ by $S_{i}$ and the other columns by $U_{i}=\left\{ 1,\dots,n\right\} \setminus S_{i}$,
V-BLAST extends $S_{i}$ according to the rule
\[
S_{i+1}=S_{i}\cup\arg\max_{j\in U_{i}}\left\Vert A_{1:m,j}-\Span{A_{1:m,U_{i}\setminus\left\{ j\right\} }}\right\Vert _{2}\;.
\]

The rationale behind V-BLAST is statistical. Under the assumption
that $b=A\mathring{x}+e$ for some admissible vector $\mathring{x}$
that we want to recover and a vector $e$ of random noise with nice
statistical properties, the Babai point recovered from the V-BLAST
ordering selects a column ordering that estimates elements of $x$
in increasing variance order. That is, V-BLAST recovers first the
symbol (element of $x$) whose continuous (real or complex) maximum-likelihood
estimate has the smallest variance. The symbol with the next smallest
continuous uncertainty (conditioned on the discrete assignment of
the first symbol) is recovered next, and so on~\cite{doi:10.1109/issse.1998.738086}. 

Surprisingly, even though V-BLAST is defined as a greedy algorithm,
it does achieve some global guarantee~\cite{doi:10.1109/issse.1998.738086}:
\[
P_{\text{V-BLAST}}=\arg\max_{P}\left\{ \min_{i}\left|R_{i,i}\right|\;\text{such that}\;AP=QR\right\} \;.
\]
That is, V-BLAST maximizes the magnitude of the smallest diagonal
element in the $R$ factor of $AP$ over all permutations. There can
obviously be more than one maximizing permutation. 

The original algorithm~\cite{doi:10.1109/issse.1998.738086} proposed
for V-BLAST performs $\Theta(mn^{3})$ operations. The algorithm computes
the Moore-Penrose pseudoinverse $A^{+}$ of $A$, finds the index
$i$ of the row of $A^{+}$ with the smallest norm, and appends $i$
to the end of the permutation. That is, column $i$ of $A$ becomes
column $n$ of $AP$. Then the algorithm zeros column $i$ in $A$,
computes the pseudoinverse again, and continues, ignoring row $i$
in the pseudoinverse when selecting the next minimal-norm row, and
so on.

This is very inefficient. Several methods that are asymptotically
more efficient have been proposed in the literature. They are described
later in this section.

\subsection{The Algorithms by Hassibi and by Chang and Paige}

Hassibi~\cite{doi:10.1109/icassp.2000.859065} and later Chang and
Paige~\cite{doi:10.1016/j.apnum.2007.01.008} independently proposed
much more efficient algorithms. The two algorithms are closely related. 

Both algorithms consider (but do not necessarily compute) the QR factorization
$A=QR$. They then construct $P$ from $R^{-1}$. Hassibi assumes
that 
\[
A=\left[\begin{array}{c}
H\\
\lambda I
\end{array}\right]
\]
and proposes to compute $R^{-1}$ directly from $A$ using a so-called
\emph{square-root} algorithm. In essence, his algorithm for computing
$R^{-1}$ starts from $(1/\lambda)I$ and updates the inverse $R$
factor using a sequence of update operations, one for each row of
$H$. Chang and Paige do not make any additional assumptions. They
propose to compute the $R$ factor using Householder reflections or
Givens rotations and then to invert $R$ explicitly. 

Both approaches for computing $R^{-1}$ require $\Theta(mn^{2})$
operations. The details of this phase are mostly irrelevant for this
article, except that we note that the $R$ factor can be computed
stably using parallel and communication-efficient algorithms~\cite{doi:10.1137/080731992}.

From this point, both algorithms continue in a similar way, but the
presentation in Hassibi's article is very terse, so we follow that
of Chang and Paige. Let $L=R^{-T}$. The diagonal elements of $L$
are the inverses of diagonal elements in $R$. The algorithm constructs
the permutation from the last (rightmost) column to the first (leftmost),
aiming to minimize the magnitude of the diagonal element of $L$ in
every step. 

We denote by $P^{(j)}$ the permutation constructed in steps $1$
through $j$ of the algorithm and by $L^{(j)}$ the transpose of the
inverse $R$ factor of $AP^{(j)}$. We initialize $P^{(0)}=I$ and
$L^{(0)}=R^{-T}$, where $R$ is the factor of $A$.

In step $j$, the algorithm finds the column of $A$ that should be
in position $n-j+1$ in $AP$. It finds 
\[
i=\arg\min_{i\le n-j+1}\left\Vert L^{(j-1)}_{1:n-j+1,i}\right\Vert _{2}\;.
\]

Now the algorithm constructs $P^{(j)}$ from $P^{(j-1)}$ by keeping
the last $j-1$ columns, setting column $n-j+1$ to 
\[
P^{(j)}_{:,n-j+1}=P^{(j-1)}_{:,i}\;,
\]
sliding columns $i+1:n-j+1$ one column to the left,
\[
P^{(j)}_{:,i:n-j}=P^{(j-1)}_{:,i+1:n-j+1}\;.
\]
The first $i-1$ columns are not modified.

The columns of $L$ undergo the same transformation,
\begin{eqnarray*}
\bar{L}^{(j)}_{:,1:i-1} & = & L^{(j-1)}_{:,1:i-1}\\
\bar{L}^{(j)}_{:,i:n-j} & = & L^{(j-1)}_{:,i+1:n-j+1}\\
\bar{L}^{(j)}_{:,n-j+1} & = & L^{(j-1)}_{:,i}\\
\bar{L}^{(j)}_{:,n-j+2:n} & = & L^{(j-1)}_{:,n-j+2:n}\;.
\end{eqnarray*}
Finally, we perform a series of rotations on pairs of rows of $\bar{L}^{(j)}$
to return this matrix to lower-triangular form $L^{(j)}$. Mathematically,
this modifies the $Q$ factor of the $QR$ factorization of $AP$,
but this is fine since the $Q$ factor is irrelevant for V-BLAST.
The nonzero structure of $\bar{L}^{(j)}$ is
\[
\setlength{\arraycolsep}{3.0pt}\bar{L}^{(j)}=\left[\begin{array}{cccccccccc}
\bullet\\
\vdots & \ddots\\
 &  & \bullet\\
 &  & \bullet &  &  &  & \bullet\\
 &  & \bullet & \bullet &  &  & \bullet\\
\vdots &  & \vdots & \vdots & \ddots &  & \vdots\\
 &  &  &  &  & \bullet & \bullet\\
 &  &  &  &  & \bullet & \bullet & \bullet\\
\vdots &  &  &  &  & \vdots & \vdots &  & \ddots\\
\bullet & \cdots & \bullet & \bullet & \cdots & \bullet & \bullet & \bullet & \cdots & \bullet
\end{array}\right]\;.
\]
Now the algorithm performs a series of rotations on pairs of adjacent
rows, starting from $i,i+1$ and going down to $n-j,n-j+1$. This
eliminates the spike in column $n-j+1$ and fills back the main diagonal
in columns $i$ through $n-j$. This gives us $L^{(j)}$ and we can
start the next step.

Hassibi also uses a unitary transformation to return the inverse $R$
factor back to triangular form after a column is moved to its final
position, but does not specify how to construct this transformation.

Let us now analyze the key performance characteristics of this algorithm
to construct $P=P^{(n)}$. In step $j$, we perform $O(n^{2})$ operations
to find $i$ (this can be reduced to $O(n)$ by maintaining an array
of squared column norms and updating it at the end of every step).
Moving columns of $L$ also requires $O(n^{2})$ operations, but this
too can be reduced to $O(n)$ using an array of pointers to columns.
The main cost in step $j$ is the sequence of Givens rotations. There
are $O(n)$ rotations and each one is applied to two rows with $O(n)$
elements, so the total cost of the Givens rotations is $O(n^{2})$.
Therefore, the total cost of this phase is $O(n^{3})$. The literature
does not contain an analysis of other performance predictors; we analyze
them in the next section.

\subsection{Algorithms Based on Downdating a Pseudoinverse}\label{subsec:pseudo-inverse-downdating}

Benesty, Huang, and Chen~\cite{doi:10.1109/tsp.2003.812897} proposed
another approach to reduce the complexity of the original V-BLAST
algorithms. Their algorithm was later improved by Luo et al.~\cite{doi:10.1109/twc.2007.05806},
Szczeciński and Massicotte~\cite{doi:10.1109/glocom.2005.1578079}
and Shang and Xia~\cite{doi:10.1109/twc.2009.081466}. All those
works share the same main ideas, which we present here. 

Consider the original V-BLAST algorithm~\cite{doi:10.1109/issse.1998.738086}.
The algorithm zeros out a column of $A$ in every iteration and permutes
the columns so that the zeroed-out column is at the end. Since we
are interested only in the permutation, we can leave the zeroed-out
column in place. We denote the matrix after iteration $i$, in which
$i$ columns have been zeroed out, by $A^{(i)}$. The original algorithm
computes the Moore-Penrose pseudoinverse of each $A^{(i)}$, in order
to identify the row with the smallest norm in $\left(A^{\left(i\right)}\right)^{+}$,
ignoring rows corresponding to already-zeroed columns in $A^{(i)}$.
The index of this row is the index of the column that is zeroed out
in iteration $i+1$.

The algorithm of Benesty et al.~\cite{doi:10.1109/tsp.2003.812897}
is based on two ideas. The first is the observation that the norms
of rows in $\left(A^{\left(i\right)}\right)^{+}$ are the square roots
of diagonal elements of the pseudoinverse of the Gram matrix of $A^{(i)}$,
\[
\left(\left(A^{\left(i\right)}\right)^{*}A^{\left(i\right)}\right)^{+}_{j,j}=\left\Vert \left(A^{\left(i\right)}\right)^{+}_{j,:}\right\Vert ^{2}_{2}\;.
\]
Denote 
\[
M^{\left(i\right)}=\left(\left(A^{\left(i\right)}\right)^{*}A^{\left(i\right)}\right)^{+}\;.
\]
Denoting the column to be eliminated in step $i+1$ by $j$, Benesty
et al.~show that 
\[
M^{\left(i+1\right)}=M^{\left(i\right)}-\frac{M^{\left(i\right)}_{:,j}\cdot\left(M^{\left(i\right)}_{:,j}\right)^{*}}{M^{\left(i\right)}_{j,j}}\ .
\]
This allows the sequence of $M^{(i)}$s to be computed using a series
of downdating operations starting from $M^{(0)}=(A^{*}A)^{-1}$. 

The different variants of this approach that have been proposed in
the literature differ in how $M^{(0)}$ is computed and in aspects
that are irrelevant to the construction of the V-BLAST permutation.
These issues are all largely irrelevant to our paper.

\subsection{A Superficial Connection to QR with Column Pivoting}

We note that the construction of the V-BLAST permutation bears a superficial
similarity to the QR factorization with column pivoting. QR with column
pivoting computes the QR factorization $AP=QR$ of a column permutation
of $A$ and constructs the permutation at the same time. The factorization
is right looking. After the first $i-1$ columns have been selected,
the algorithm selects the next column for elimination in a way that
maximizes $|R_{i,i}|$. 

QR with column pivoting is a traditional strategy to compute a rank-revealing
QR factorization; today, there are additional strategies that often
deliver higher performance. 

We exploit this observation in the design of new algorithms for V-BLAST.

\section{Complexity Analysis of Existing Algorithms}\label{sec:Complexity-Analysis-of-Existing-Algorithms}

The original V-BLAST algorithm~\cite{doi:10.1109/issse.1998.738086}
is asymptotically expensive. Computing the pseudoinverses of a sequence
of $n$ matrices, each time from scratch, requires performing $\Theta(mn^{3})$
arithmetic operations, assuming Strassen-like methods are not used. 

Both Hassibi~\cite{doi:10.1109/icassp.2000.859065} and Chang and
Paige~\cite{doi:10.1016/j.apnum.2007.01.008} proposed algorithms
that require only $\Theta(mn^{2})$ operations, significantly improving
the asymptotic performance. Most of the other algorithms in the wireless-communication
literature focus on reducing the required number of arithmetic operations
by a constant factor.

However, the actual performance of an algorithm on a modern computer
depends not only on the number of arithmetic operations that the algorithm
performs, but also on the ability of the algorithm to exploit multiple
processors or arithmetic units and on the pattern of data movement
in the algorithm. These aspects of an algorithm can have a much stronger
influence on its running time than the total number of arithmetic
operations. 

\subsection{Simple Models of Parallelism and Data Reuse}

We analyze parallelism in terms of the critical path of algorithms.
That is, we characterize the available parallelism in an algorithm
using the longest chain of dependencies in the algorithm, called its
critical path. The length of the critical path, called span (or just
critical path), bounds the running time independently of how many
processors are available and how fast data movement is. The shorter
the critical path, the more likely it is that the algorithm can exploit
multiple processing units. In particular, in many matrix algorithms
the span, denoted $T_{\infty}$, is a good predictor of performance.
More specifically, under fairly conservative assumptions on the algorithm
but using generous assumptions on the memory system, $p$ processors
can execute an algorithm that performs $T_{1}$ operations in a little
over $T_{\infty}+T_{1}/p$ steps~\cite{doi:10.1006/jpdc.1996.0107}.

We analyze the cost of data movement using a sequential two-level
memory hierarchy model, known formally as the Hong-Kung pebble game~\cite{doi:10.1145/800076.802486,doi:10.1090/dimacs/050/09}.
Formally, we assume that the computer has a fast memory (a cache)
with $M$ words and a slow memory of infinite size. Processors can
only operate on data in fast memory and can only store results to
fast memory. We also assume that any subset of the algorithm's data
structure can reside concurrently in the cache and that cache evictions
are performed optimally. We characterize the performance of an algorithm
using the number of data transfers it must perform between slow and
fast memory.

For most of the common matrix factorizations, such as the LU factorization
with partial pivoting, the Cholesky factorization, and the QR factorization,
there are practical and backward stable algorithms with critical path
$\Theta(n)$ and cache-efficient algorithms that move $\Theta(T_{1}/\sqrt{M})$
words between fast and slow memory~\cite{doi:10.1137/090769156}.
These data-movement upper bounds are asymptotically optimal. There
are factorization algorithms with polylogarithmic span, but they are
numerically unstable.

We do not address in this article more complex theoretical models
of data movement (or parallelism), such as distributed-memory models,
communication-avoiding algorithms, and so on.

\subsection{Parallelism and Cache Misses in the Algorithms of Hassibi and of
Chang and Paige}\label{subsec:parallelism-in-chang-paige}

Both algorithms first compute the inverse $R^{-1}$ of the R factor
of $A=QR$. If $R$ is computed using Householder reflections, the
critical path has length $\Theta(n)$ and there are cache-efficient
variants~\cite{doi:10.1137/13092157X}. The inverse of $R$ can be
computed by substitution with critical path $\Theta(n)$ and optimal
data reuse. There are also triangular-inversion algorithms with logarithmic
critical path and good data reuse. There are high-quality implementations
of these algorithms. We consider these preprocessing steps as resolved
in terms of parallelism and data reuse. We do note that Hassibi and
some of the follow-up algorithms proposed to compute $R^{-1}$ in
other ways, which are both less numerically stable and seem to have
much worse data reuse ratios. 

The length of the critical path in the algorithm of Chang and Paige
is $\Theta(n^{2})$, asymptotically much longer than the critical
path in most matrix factorizations. The relevant details are missing
from Hassibi's algorithm, but it appears to suggest the same algorithm.

More specifically, the algorithm performs $\Theta(n^{2})$ Givens
rotations. Each rotation can be applied with constant critical path,
giving the upper bound. As far as we can see, the Givens rotations
must be performed sequentially, and any such implementation has critical
path $\Theta(n^{2})$.

This ordering of the Givens rotations implies that in most of the
$n-1$ steps of the algorithm, it might transform a constant factor
of the elements of $L$, so the number of cache misses with a memory
of $M\leq n^{2}/8$ words is $\Omega(n^{3})$. 

\subsection{Parallelism and Cache Misses in the Algorithms that Downdate a Pseudoinverse}\label{subsec:parallelism-in-downdating}

We again ignore the preprocessing phase that computes $(A^{*}A)^{-1}$,
because it can be performed in several ways using algorithms with
linear critical path and with good data reuse ratios. These include
computing $R^{-1}$ and then multiplying $(A^{*}A)^{-1}=(R^{*}Q^{*}QR)^{-1}=(R^{*}R)^{-1}=R^{-1}R^{-*}$,
or even the SVD. We again note that most of the variants in the wireless-communication
literature proposed other ways of computing the initial pseudoinverse
$(A^{*}A)^{-1}$, which appear to suffer from suboptimal data reuse
ratios. We again consider the performance aspects of this preprocessing
step to be resolved.

The algorithm by Benesty et al.~\cite{doi:10.1109/tsp.2003.812897}
performs a sequence of $n$ downdating operations on a square symmetric
(Hermitian) matrix. The downdating operations are simple; they amount
to simple rank-1 updates with constant critical path. After each rank-1
update, the algorithm finds an admissible minimum diagonal element;
this operation has a logarithmic critical path. Therefore, the length
of the critical path is $\Theta(n\log n)$. The data reuse ratio for
$M\leq n^{2}/8$ is constant, so the number of cache misses is $\Omega(n^{3})$;
this aspect of these algorithms is poor.

\section{Constructing the Permutation Using Orthogonal Transformations and
an Almost Linear Span}\label{sec:VBLAST-via-Householder}

We now present a variant of the Hassibi/Chang and Paige algorithm
with an almost linear span. Like their algorithms, it uses only unitary
transformations on $R^{-T}$, so it is less likely to suffer from
numerical problems than the pseudo-inverse downdating approaches.

The idea is simple. We start from $L=R^{-T}$, compute the 2-norms
of its columns, and exchange the column with the smallest norm with
the last. Instead of eliminating the spike in column $n$ with Givens
rotations, as Chang and Paige do, we use a Householder reflection.
This fills the elements in the upper triangle in rows and columns
$1:n-1$. We do not attempt to bring the matrix back into lower triangular
form. We update the column norms to account for the exchange and to
exclude the elements in the last row. Step~1 is now complete. We
search for the shortest column $j'$ among $1:n-1$, exchange it with
column $n-1$, and eliminate the spike in column $n-1$. Note that
the Householder reflection does not affect row $n$, so the nonzero
structure of the last column is preserved.

We then repeat, keeping the invariant that the Householder reflection
in step $i$ does not affect rows $n-i+2:n$. Therefore, after step
$i$, the last $i$ columns are a lower triangular matrix, 
\[
\setlength{\arraycolsep}{3.0pt}\bar{L}^{(i)}=\left[\begin{array}{ccccccccc}
\bullet & \cdots & \bullet & \cdots & \bullet & \bullet\\
\bullet &  & \bullet &  & \bullet & \bullet\\
\vdots &  & \vdots &  &  & \vdots\\
\bullet &  & \bullet &  & \bullet & \bullet\\
\bullet &  & \bullet &  & \bullet & \bullet\\
\bullet &  & \bullet &  & \bullet & \bullet & \bullet\\
\vdots &  & \vdots &  & \vdots & \vdots &  & \ddots\\
\bullet & \cdots & \bullet & \cdots & \bullet & \bullet & \bullet & \cdots & \bullet
\end{array}\right]\;.
\]
At this point, the algorithm exchanges the column with the smallest
norm (excluding the bottom $i$ elements), shown in blue below, with
column $n-i$, shown in red.
\[
\setlength{\arraycolsep}{3.0pt}\bar{L}^{(i)}=\left[\begin{array}{ccccccccc}
\bullet & \cdots & \mathbin{\color{blue}\bullet} & \cdots & \bullet & \mathbin{\color{red}\bullet}\\
\bullet &  & \mathbin{\color{blue}\bullet} &  & \bullet & \mathbin{\color{red}\bullet}\\
\vdots &  & {\color{blue}\vdots} &  &  & {\color{red}\vdots}\\
\bullet &  & \mathbin{\color{blue}\bullet} &  & \bullet & \mathbin{\color{red}\bullet}\\
\bullet &  & \mathbin{\color{blue}\bullet} &  & \bullet & \mathbin{\color{red}\bullet}\\
\bullet &  & \mathbin{\color{blue}\bullet} &  & \bullet & \mathbin{\color{red}\bullet} & \bullet\\
\vdots &  & {\color{blue}\vdots} &  & \vdots & {\color{red}\vdots} &  & \ddots\\
\bullet & \cdots & \mathbin{\color{blue}\bullet} & \cdots & \bullet & \mathbin{\color{red}\bullet} & \bullet & \cdots & \bullet
\end{array}\right]\quad\longrightarrow\quad\left[\begin{array}{ccccccccc}
\bullet & \cdots & \mathbin{\color{red}\bullet} & \cdots & \bullet & \mathbin{\color{blue}\bullet}\\
\bullet &  & \mathbin{\color{red}\bullet} &  & \bullet & \mathbin{\color{blue}\bullet}\\
\vdots &  & {\color{red}\vdots} &  &  & {\color{blue}\vdots}\\
\bullet &  & \mathbin{\color{red}\bullet} &  & \bullet & \mathbin{\color{blue}\bullet}\\
\bullet &  & \mathbin{\color{red}\bullet} &  & \bullet & \mathbin{\color{blue}\bullet}\\
\bullet &  & \mathbin{\color{red}\bullet} &  & \bullet & \mathbin{\color{blue}\bullet} & \bullet\\
\vdots &  & {\color{red}\vdots} &  & \vdots & {\color{blue}\vdots} &  & \ddots\\
\bullet & \cdots & \mathbin{\color{red}\bullet} & \cdots & \bullet & \mathbin{\color{blue}\bullet} & \bullet & \cdots & \bullet
\end{array}\right]\;.
\]
A Householder reflection on the rows shown in grey extends the lower-triangular
structure by one more column,
\[
\setlength{\arraycolsep}{3.0pt}\left[\begin{array}{ccccccccc}
\mathbin{\color{lightgray}\bullet} & {\color{lightgray}\cdots} & \mathbin{\color{lightgray}\bullet} & {\color{lightgray}\cdots} & \mathbin{\color{lightgray}\bullet} & \mathbin{\color{lightgray}\bullet} & {\color{black}} & {\color{black}} & {\color{black}}\\
\mathbin{\color{lightgray}\bullet} & {\color{lightgray}{\color{black}}} & \mathbin{\color{lightgray}\bullet} & {\color{lightgray}{\color{black}}} & \mathbin{\color{lightgray}\bullet} & \mathbin{\color{lightgray}\bullet} & {\color{black}} & {\color{black}} & {\color{black}}\\
{\color{lightgray}\vdots} & {\color{lightgray}{\color{black}}} & {\color{lightgray}\vdots} & {\color{lightgray}{\color{black}}} & {\color{lightgray}{\color{black}}} & {\color{lightgray}\vdots} & {\color{black}} & {\color{black}} & {\color{black}}\\
\mathbin{\color{lightgray}\bullet} & {\color{lightgray}{\color{black}}} & \mathbin{\color{lightgray}\bullet} & {\color{lightgray}{\color{black}}} & \mathbin{\color{lightgray}\bullet} & \mathbin{\color{lightgray}\bullet} & {\color{black}} & {\color{black}} & {\color{black}}\\
\mathbin{\color{lightgray}\bullet} & {\color{lightgray}{\color{black}}} & \mathbin{\color{lightgray}\bullet} & {\color{lightgray}{\color{black}}} & \mathbin{\color{lightgray}\bullet} & \mathbin{\color{lightgray}\bullet} & {\color{black}} & {\color{black}} & {\color{black}}\\
\mathbin{\color{black}\bullet} & {\color{black}} & \mathbin{\color{black}\bullet} & {\color{black}} & \mathbin{\color{black}\bullet} & \mathbin{\color{black}\bullet} & \mathbin{\color{black}\bullet} & {\color{black}} & {\color{black}}\\
{\color{black}\vdots} & {\color{black}} & {\color{black}\vdots} & {\color{black}} & {\color{black}\vdots} & {\color{black}\vdots} & {\color{black}} & {\color{black}\ddots} & {\color{black}}\\
\mathbin{\color{black}\bullet} & {\color{black}\cdots} & \mathbin{\color{black}\bullet} & {\color{black}\cdots} & \mathbin{\color{black}\bullet} & \mathbin{\color{black}\bullet} & \mathbin{\color{black}\bullet} & {\color{black}\cdots} & \mathbin{\color{black}\bullet}
\end{array}\right]\quad\longrightarrow\quad\left[\begin{array}{ccccccccc}
\mathbin{\color{lightgray}\bullet} & {\color{lightgray}\cdots} & \mathbin{\color{lightgray}\bullet} & {\color{lightgray}\cdots} & \mathbin{\color{lightgray}\bullet} &  & {\color{black}} & {\color{black}} & {\color{black}}\\
\mathbin{\color{lightgray}\bullet} & {\color{lightgray}{\color{black}}} & \mathbin{\color{lightgray}\bullet} & {\color{lightgray}{\color{black}}} & \mathbin{\color{lightgray}\bullet} &  & {\color{black}} & {\color{black}} & {\color{black}}\\
{\color{lightgray}\vdots} & {\color{lightgray}{\color{black}}} & {\color{lightgray}\vdots} & {\color{lightgray}{\color{black}}} & {\color{lightgray}{\color{black}}} &  & {\color{black}} & {\color{black}} & {\color{black}}\\
\mathbin{\color{lightgray}\bullet} & {\color{lightgray}{\color{black}}} & \mathbin{\color{lightgray}\bullet} & {\color{lightgray}{\color{black}}} & \mathbin{\color{lightgray}\bullet} &  & {\color{black}} & {\color{black}} & {\color{black}}\\
\mathbin{\color{lightgray}\bullet} & {\color{lightgray}{\color{black}}} & \mathbin{\color{lightgray}\bullet} & {\color{lightgray}{\color{black}}} & \mathbin{\color{lightgray}\bullet} & \mathbin{\color{lightgray}\bullet} & {\color{black}} & {\color{black}} & {\color{black}}\\
\mathbin{\color{black}\bullet} & {\color{black}} & \mathbin{\color{black}\bullet} & {\color{black}} & \mathbin{\color{black}\bullet} & \mathbin{\color{black}\bullet} & \mathbin{\color{black}\bullet} & {\color{black}} & {\color{black}}\\
{\color{black}\vdots} & {\color{black}} & {\color{black}\vdots} & {\color{black}} & {\color{black}\vdots} & {\color{black}\vdots} & {\color{black}} & {\color{black}\ddots} & {\color{black}}\\
\mathbin{\color{black}\bullet} & {\color{black}\cdots} & \mathbin{\color{black}\bullet} & {\color{black}\cdots} & \mathbin{\color{black}\bullet} & \mathbin{\color{black}\bullet} & \mathbin{\color{black}\bullet} & {\color{black}\cdots} & \mathbin{\color{black}\bullet}
\end{array}\right]\;.
\]

We note that this algorithm is an exact mirror of QR with column pivoting:
\begin{itemize}
\item We work with a lower-triangular matrix rather than an upper triangular
one,
\item we construct the permutation from the end (the right) instead of constructing
it from the beginning (the left),
\item and we choose the column with the smallest norm (in the rows that
are not yet part of the triangular structure) instead of the column
with the largest norm.
\end{itemize}

\subsection{Analyses of Parallelism and Cache Misses}

We ignore the computation of $L=R^{-T}$, since as noted in Subsection~\textcolor{blue}{\ref{subsec:parallelism-in-chang-paige}}
it can be done with a short span and in a cache-efficient manner. 

We start with an analysis of parallelism in the Householder-based
algorithm described above. The algorithm has $n-1$ sequential steps.
In each step it selects one column. In each step:
\begin{itemize}
\item the algorithm computes the norms of all remaining columns,
\item computes a Householder reflection of the column with the smallest
norm,
\item and applies the reflection to the rest of the remaining columns.
\end{itemize}
The span of these phases is $\Theta(\log n)$. Therefore, the total
span of the algorithm is $\Theta(n\log n)$. This is a significant
improvement over the algorithm of Chang and Paige, and the same asymptotic
bound as the downdating approach.

Unfortunately, this algorithm is not cache efficient. Updating the
remaining columns in each step requires accessing the remaining submatrix.
The number of operations on each of its elements is a small constant,
so with a memory of $M\leq n^{2}/8$ words the number of cache misses
is $\Omega(n^{3})$.

\section{Approximating V-BLAST with Random Projections}\label{sec:R-V-BLAST}

The algorithm proposed in Section~\ref{sec:VBLAST-via-Householder}
is asymptotically efficient in terms of the number of operations and
it has a reasonably short span. However, its data-access pattern is
poor, since it scans the entire reduced submatrix during the processing
of each column. QR with column pivoting has similar behavior. Attempts
to design rank-revealing QR factorizations with better data-access
patterns led to at least two families of algorithms. We show in this
section that ideas from one of them, based on random projections~\cite{doi:10.1137/15M1044680},
are applicable to V-BLAST. (The other family is based on tournaments~\cite{doi:10.1137/13092157X};
we have not found a way to apply this idea to V-BLAST.)

Our algorithm starts, like that of Chang and Paige and that of Section~\ref{sec:VBLAST-via-Householder},
from $L=R^{-T}$, where $A=QR$. The algorithm is blocked (or partitioned),
in the sense that a scan of the entire reduced submatrix allows us
to select the next block of $b$ consecutive columns in $AP$, not
just one. The algorithm is randomized, so the columns in the selected
block are not necessarily the $b$ columns that V-BLAST would select
next. 

We start by constructing a random $(b+p)$-by-$n$ sketch of the $n$-by-$n$
matrix $L$. The sketch is constructed by multiplying $L$ from the
left by a $(b+p)$-by-$n$ random matrix $G$, which has random-projection
properties (i.e, with high probability it preserves norms as well
as certain relationships between the columns inside subspaces, an
Oblivious Subspace Embedding~\cite{doi:10.1561/0400000060}). Our
theory (Section~\ref{sec:theoretical}) analyzes $G$ with i.i.d.
standard normal entries. The results can be extended to other types
of Oblivious Subspace Embeddings (OSEs), with different bounds.

We now run $b$ steps of the Householder-based algorithm from Section~\ref{sec:VBLAST-via-Householder}
on $GL$. More specifically, the first column that we choose is the
one with the minimal Euclidean norm in $GL$. We then eliminate the
top $b+p-1$ elements from this column using a Householder reflection,
and continue with the top $b+p-1$ rows of the reduced matrix. We
denote the permutation that moves the $b$ selected columns to the
end, in order, by $P^{(1:b)}$.

We now reorder $L$ accordingly, replacing $L$ by $LP^{(1:b)}$.
Next, we lower triangularize the rightmost $b$ columns in $LP^{(1:b)}$
(the columns that we selected using the sketch) using $b$ Householder
reflections. These $b$ reflections are now applied in blocks of $\Theta(b)$
columns to the rest of $LP^{(1:b)}$, using the WY representation~\cite{doi:10.1137/0910005}. 

This process leaves us with a reduced matrix of the form
\[
\setlength{\arraycolsep}{3.0pt}\left[\begin{array}{ccccccccc}
\mathbin{\color{lightgray}\bullet} & {\color{lightgray}\cdots} & \mathbin{\color{lightgray}\bullet} & {\color{lightgray}\cdots} & \mathbin{\color{lightgray}\bullet} & \mathbin{\color{lightgray}\bullet} & {\color{black}} & {\color{black}} & {\color{black}}\\
\mathbin{\color{lightgray}\bullet} & {\color{lightgray}{\color{black}}} & \mathbin{\color{lightgray}\bullet} & {\color{lightgray}{\color{black}}} & \mathbin{\color{lightgray}\bullet} & \mathbin{\color{lightgray}\bullet} & {\color{black}} & {\color{black}} & {\color{black}}\\
{\color{lightgray}\vdots} & {\color{lightgray}{\color{black}}} & {\color{lightgray}\vdots} & {\color{lightgray}{\color{black}}} & {\color{lightgray}{\color{black}}} & {\color{lightgray}\vdots} & {\color{black}} & {\color{black}} & {\color{black}}\\
\mathbin{\color{lightgray}\bullet} & {\color{lightgray}{\color{black}}} & \mathbin{\color{lightgray}\bullet} & {\color{lightgray}{\color{black}}} & \mathbin{\color{lightgray}\bullet} & \mathbin{\color{lightgray}\bullet} & {\color{black}} & {\color{black}} & {\color{black}}\\
\mathbin{\color{lightgray}\bullet} & {\color{lightgray}{\color{black}}} & \mathbin{\color{lightgray}\bullet} & {\color{lightgray}{\color{black}}} & \mathbin{\color{lightgray}\bullet} & \mathbin{\color{lightgray}\bullet} & {\color{black}} & {\color{black}} & {\color{black}}\\
\mathbin{\color{black}\bullet} & {\color{black}} & \mathbin{\color{black}\bullet} & {\color{black}} & \mathbin{\color{black}\bullet} & \mathbin{\color{black}\bullet} & \mathbin{\color{black}\bullet} & {\color{black}} & {\color{black}}\\
{\color{black}\vdots} & {\color{black}} & {\color{black}\vdots} & {\color{black}} & {\color{black}\vdots} & {\color{black}\vdots} & {\color{black}} & {\color{black}\ddots} & {\color{black}}\\
\mathbin{\color{black}\bullet} & {\color{black}\cdots} & \mathbin{\color{black}\bullet} & {\color{black}\cdots} & \mathbin{\color{black}\bullet} & \mathbin{\color{black}\bullet} & \mathbin{\color{black}\bullet} & {\color{black}\cdots} & \mathbin{\color{black}\bullet}
\end{array}\right]
\]
where the triangular part on the right contains $b$ columns. The
algorithm constructs a new sketch, using an independently sampled
$G$, of the rows and columns that are not part of the already-triangularized
part of the matrix (the elements shown in gray above), uses it to
select the next $b$ columns, and so on.

The distribution of sketch matrices that the algorithm uses maps a
given input matrix $L$ (or the original $A$) into a distribution
of permutations. We show below, using both theoretical analyses and
experiments, that the probability of permutations that are almost
as effective as V-BLAST is high.

\section{Theoretical Analyses of the Random-Projections Algorithm}\label{sec:theoretical}

We now analyze the algorithm presented in Section\ \ref{sec:R-V-BLAST}.
We prove that with high probability, the permutation returned by the
algorithm has two properties that are very similar to two important
optimality properties of the V-BLAST permutation. 

The first property is a probabilistic version of V-BLAST's greedy
column-selection rule: in every iteration, V-BLAST selects the column
with the largest distance to the span of other columns not yet selected~\cite{doi:10.1016/j.apnum.2007.01.008}. 
\begin{lemma}
\label{lem:greedy-almost-optimal}Let $0<\delta<\nicefrac{1}{2}$
and $0<\varepsilon<1$. Suppose our algorithm uses projection dimension

\[
b+p\geq C\left(\frac{b\log n+\log\frac{1}{\delta}}{\varepsilon^{2}}\right)
\]
for some absolute constant $C$. Let $P$ be the permutation returned
by the algorithm described in Section~\ref{sec:R-V-BLAST} (note
that this is a random permutation). Let $U_{t}$ be the set of indices
corresponding to the first $n-t$ columns of $AP$ (that is, the indices
not yet chosen at the start of iteration $t$ of the algorithm, with
iteration counter starting from $0$). Let $c_{t+1}$ be the index
of the $t+1$ column our algorithm chooses (chosen in iteration $t$),
meaning $A_{1:m,c_{t+1}}=\left(AP\right)_{1:m,n-t}$.

Then, with probability of at least $1-\delta$, for every $0\le t\le n-1$
we have

\begin{equation}
\left\Vert A_{1:m,c_{t+1}}-\Span{A_{1:m,U_{t}\setminus\left\{ c_{t+1}\right\} }}\right\Vert _{2}\ge\frac{1-\varepsilon}{1+\varepsilon}\max_{j\in U_{t}}\left\Vert A_{1:m,j}-\Span{A_{1:m,U_{t}\setminus\left\{ j\right\} }}\right\Vert _{2}\label{eq:lem1-property}
\end{equation}
\end{lemma}

This means while our algorithm may not choose the best available column
in each iteration as V-BLAST does, it chooses an almost optimal column:
each column it chooses has distance close to the optimal available
distance, subject to previous choices made by the algorithm. Note
that replacing $\frac{1-\varepsilon}{1+\varepsilon}$ by $1$ in the
above property means the algorithm chooses the column furthest away
from the span of the columns not yet chosen, which is exactly what
V-BLAST does~\cite{doi:10.1016/j.apnum.2007.01.008}.

To prove Lemma~\ref{lem:greedy-almost-optimal}, we need the following
auxiliary lemma. It states that the projection roughly preserves the
columns of $L$ (recall that $L=R^{-T}$, where $A=QR$). Before presenting
the lemma, we need to introduce some notation. Let $l_{i}$ be the
$i$ column of $L$ ($l_{i}=L_{1:n,i}$). Let $S_{t}=\left[n\right]\setminus U_{t}$
denote the set of indices of the first $t$ columns selected by the
algorithm, with $t=0,\dots,n$. Let $l^{(t)}_{i}$ be the orthogonal
projection of $l_{i}$ onto the orthogonal complement of $\Span{l_{j}|j\in S_{t}}$.
Note that V-BLAST chooses at iteration $t$ the column that minimizes
$\Vert l^{(t)}_{i}\Vert{}_{2}$ over $i\in U_{t}$~\cite{doi:10.1016/j.apnum.2007.01.008}.
Our algorithm does not compute $l^{(t)}_{i}$ or its norm explicitly,
except when $t$ is a multiple of $b$, as described below. To approximate
$\Vert l^{(t)}_{i}\Vert{}_{2}$, our algorithm relies on sketches
applied to the reduced matrix. After $t$ columns have been selected,
let $w=b\cdot\lfloor\frac{t}{b}\rfloor$ be the index of the first
step of the block, which is also the number of columns selected in
all fully completed prior blocks (we selected $b$ columns in each
of these blocks). 

Denote by $V_{t}=S_{t}\setminus S_{w}$ the set of indices of the
$t\mod b$ columns chosen in the current block at the start of iteration
$t$. Let $H_{w}$ be the $n\times n$ unitary matrix representing
the accumulated Householder reflections applied on $L$ so far (note
that Householder reflectors are applied to $L$ only on block boundaries;
this is why we define $H_{w}$ only for $w$ that is a multiple of
$b$). Denote the $n-w$ dimensional vector $r^{(w)}_{i}=(H_{w}l_{i})_{1:(n-w)}$,
i.e., column $i$ in the reduced matrix after $w$ column selections.
When $H_{w}$ is applied to $l_{i}$, the bottom $w$ coordinates
of $H_{w}l_{i}$ have the same norm as the projection of $l_{i}$
onto $\Span{l_{j}|j\in S_{w}}$, since the $w$ right-most columns
of $H_{w}(LP)$ are triangular. Since $H_{w}$ is unitary, this means
that the top $n-w$ elements of $H_{w}l_{i}$ have the same norm as
the orthogonal projection of $l_{i}$ onto the orthogonal complement
of $\Span{l_{j}|j\in S_{w}},$ i.e., $\lVert r^{(w)}_{i}\rVert_{2}=\lVert l^{(w)}_{i}\rVert_{2}$
(see Appendix~(\ref{app:aux}) for a formal proof\footnote{Since this is a highly technical proof, which doesn't contribute to
the flow of the main proof, we defer it to the appendix.}).

Since $r^{(w)}_{i}$ is computed explicitly (after applying the $b$
blocked Householder reflectors of each step), we see that we indeed
explicitly compute $\Vert l^{(t)}_{i}\Vert{}_{2}$ when $t$ is a
multiple of $b$, as mentioned above. The algorithm samples an independent
sketch matrix $G_{w}$ of size $(b+p)\times(n-w)$ for the block starting
at iteration $w$. We denote $\tilde{l}^{(w)}_{i}=G_{w}r^{(w)}_{i}$,
and define $\tilde{l}^{(t)}_{i}$ as the orthogonal projection of
$\tilde{l}^{(w)}_{i}$ onto the orthogonal complement of $\Span{\tilde{l}^{(w)}_{j}|j\in V_{t}}$.
Using the same reasoning as above, our algorithm chooses the $\left(t+1\right)-th$
column by choosing the column with the minimal $\Vert\tilde{l}^{(t)}_{i}\Vert{}_{2}$,
which it explicitly computes in each step.

We can now state the auxiliary lemma. It states that our algorithm
can estimate $\Vert l^{(t)}_{i}\Vert{}_{2}$ (which it does not compute)
using $\Vert\tilde{l}^{(t)}_{i}\Vert{}_{2}$ (which it does compute).
\begin{lemma}
\label{lem:block-norm-preserve}

Let $0<\delta<\nicefrac{1}{2}$ and $0<\varepsilon<1$. Suppose our
algorithm uses projection dimension

\[
b+p\ge C\left(\frac{b\log n+\log\frac{1}{\delta}}{\varepsilon^{2}}\right)
\]
for some absolute constant C.

Then with probability at least $1-\delta$, for every feasible $i$
and $t$,

\[
\left(1-\varepsilon\right)\left\Vert l^{(t)}_{i}\right\Vert _{2}\le\frac{1}{\sqrt{b+p}}\left\Vert \tilde{l}^{(t)}_{i}\right\Vert _{2}\le\left(1+\varepsilon\right)\left\Vert l^{(t)}_{i}\right\Vert _{2}
\]
\end{lemma}

\begin{proof}
We note that $l^{(t)}_{i}$ is $l_{i}$ after subtracting some linear
combination of the first $t$ columns chosen. Moreover, we can express
$l^{(t)}_{i}$ as a linear combination of $l^{(w)}_{j}$ for some
$j$'s: it is $l^{(w)}_{i}$ after subtracting a linear combination
of $l^{(w)}_{j}$ for $j\in V_{t}$ . We also have, by definition,
that $\tilde{l}^{(t)}_{i}$ is a linear combination of $\tilde{l}^{(w)}_{j}$
for some $j\in V_{t}\cup\left\{ i\right\} $. It is well known that
if $G$ is a $k$-by-$n$ random matrix with i.i.d. standard normal
entries, then $G/\sqrt{k}$ is an Oblivious Subspace Embedding if
$k$ is large enough. That is, for every fixed $d$-dimensional subspace,
with probability at least $1-\varDelta$, we have for every vector
$x$ in this subspace
\[
\left(1-\varepsilon\right)\lVert x\rVert_{2}\le\frac{1}{\sqrt{k}}\lVert Gx\rVert_{2}\le\left(1+\varepsilon\right)\lVert x\rVert_{2}\ .
\]
The size of $k$ to get this guarantee is $k\geq C_{0}\left(d+\log\left(\nicefrac{1}{\varDelta}\right)\right)\cdot\varepsilon^{-2}$
for some absolute constant $C_{0}$~\cite[Theorem 6]{doi:10.1561/0400000060}.

For each block $w$, condition on all randomness from the preceding
blocks. The vectors $r^{(w)}_{i}$ are then fixed, and the new sketch
$G_{w}$ is independent of this history. The following bound on the
failure probability for a fixed block holds conditionally on every
such history, and therefore also unconditionally. We take $\varDelta$
to be small enough such that the above norm preservation holds simultaneously
for all subspaces spanned by $b$ vectors from $\left\{ r^{(w)}_{1},\dots,r^{(w)}_{n}\right\} $
for any $w$. For a fixed $w$, there are at most ${n \choose b}$
such subspaces, so we can bound the number of subspaces by

\[
{n \choose b}\le\left(\frac{en}{b}\right)^{b}\ .
\]
Thus, if we choose 
\[
\varDelta=\frac{\delta}{n\cdot\left(\frac{en}{b}\right)^{b}}
\]
then for every fixed $w$, norm preservation holds for all subspaces
spanned by $b$ vectors from $\left\{ r^{(w)}_{1},\dots,r^{(w)}_{n}\right\} $
with probability at least $1-\delta/n$. Observing that there are
$\lceil\frac{n}{b}\rceil\leq n$ blocks, we now use union bound to
see that the guarantee holds for all blocks and subspaces with probability
at least $1-\delta$. Thus, the projection dimension needs to be

\[
\begin{aligned}b+p\geq C_{0}\left(\frac{b+\log\left(\frac{1}{\varDelta}\right)}{\varepsilon^{2}}\right) & =C_{0}\left(\frac{b+\log\left(\frac{n\cdot\left(\frac{en}{b}\right)^{b}}{\delta}\right)}{\varepsilon^{2}}\right)\\
 & =O\left(\frac{b\log\left(n\right)+\log\left(\frac{1}{\delta}\right)}{\varepsilon^{2}}\right)
\end{aligned}
\]
Thus, we have proved that for a sufficiently large value of $b+p$,
$G_{w}/\sqrt{b+p}$ preserves norms for all subspaces spanned by $b$
vectors from $\left\{ r^{(w)}_{1},\dots,r^{(w)}_{n}\right\} $ for
every $w$.

Denote by $q^{(t)}_{i}$ the orthogonal projection of $r^{(w)}_{i}$
into $\Span{r^{(w)}_{j}|j\in V_{t}}$. Since $r^{(w)}_{i}=(H_{w}l_{i})_{1:n-w}$
and $H_{w}$ is orthogonal, we have $\ensuremath{\left\Vert l^{(t)}_{i}\right\Vert _{2}=\left\Vert r^{(w)}_{i}-q^{(t)}_{i}\right\Vert _{2}}$
(see Appendix~\ref{app:aux} for a more formal proof). Denote by
$G_{w}\cdot\widetilde{q}^{(t)}_{i}$ the orthogonal projection of
$\widetilde{l}^{(w)}_{i}$ into $\Span{\widetilde{l}^{(w)}_{j}|j\in V_{t}}$
(since the vectors $\widetilde{l}^{(w)}_{j}$ for $j\in V_{t}$ are
in the range of $G_{w}$ then so is the projection). So $q^{(t)}_{i}$
is by definition inside $\Span{r^{(w)}_{j}|j\in V_{t}}$ , and we
can choose $\widetilde{q}^{(t)}_{i}$ to lie in the same span.

Since $q^{(t)}_{i}$ is the closest vector to $r^{(w)}_{i}$ in $\Span{r^{(w)}_{j}|j\in V_{t}}$,
we have

\[
\ensuremath{\left\Vert l^{(t)}_{i}\right\Vert _{2}=\left\Vert r^{(w)}_{i}-q^{(t)}_{i}\right\Vert _{2}}\le\left\Vert r^{(w)}_{i}-\widetilde{q}^{(t)}_{i}\right\Vert _{2}~.
\]
Since $G_{w}\cdot\widetilde{q}^{(t)}_{i}$ is the closest vector to
$\widetilde{l}^{(w)}_{i}$ in $\Span{\widetilde{l}^{(w)}_{j}|j\in V_{t}}$
and $\tilde{l}^{(w)}_{j}=G_{w}\cdot r^{(w)}_{j}$, we have

\[
\left\Vert \widetilde{l}^{(t)}_{i}\right\Vert _{2}=\left\Vert \widetilde{l}^{(w)}_{i}-G_{w}\cdot\widetilde{q}^{(t)}_{i}\right\Vert _{2}\le\left\Vert \widetilde{l}^{(w)}_{i}-G_{w}\cdot q^{(t)}_{i}\right\Vert _{2}=\left\Vert G_{w}\cdot r^{(w)}_{i}-G_{w}\cdot q^{(t)}_{i}\right\Vert _{2}~.
\]

To conclude the proof, we use the last two inequalities with the norm
preservation of all subspaces spanned by columns of all blocks: if
every $G_{w}/\sqrt{b+p}$ preserves norms of vectors in spans of the
form $\{r^{(w)}_{i}\}\cup\{r^{(w)}_{j}\,:\,j\in V_{t}\}$ up to a
factor of $1\pm\varepsilon$ (which we showed above happens with probability
at least $1-\delta$, since these subspaces are spanned by at most
$b$ vectors from $\left\{ r^{(w)}_{1},\dots,r^{(w)}_{n}\right\} $),
we have

\[
\begin{aligned}\left(1-\varepsilon\right)\left\Vert l^{(t)}_{i}\right\Vert _{2} & \le\left(1-\varepsilon\right)\left\Vert r^{(w)}_{i}-\widetilde{q}^{(t)}_{i}\right\Vert _{2}\\
 & \le\frac{1}{\sqrt{b+p}}\left\Vert G_{w}\cdot\left(r^{(w)}_{i}-\widetilde{q}^{(t)}_{i}\right)\right\Vert _{2}\\
 & =\frac{1}{\sqrt{b+p}}\left\Vert \widetilde{l}^{(w)}_{i}-G_{w}\cdot\widetilde{q}^{(t)}_{i}\right\Vert _{2}\\
 & =\frac{1}{\sqrt{b+p}}\left\Vert \tilde{l}^{(t)}_{i}\right\Vert _{2}
\end{aligned}
\]
and also

\[
\begin{aligned}\frac{1}{\sqrt{b+p}}\left\Vert \tilde{l}^{(t)}_{i}\right\Vert _{2} & \le\frac{1}{\sqrt{b+p}}\left\Vert G_{w}\cdot r^{(w)}_{i}-G_{w}\cdot q^{(t)}_{i}\right\Vert _{2}\\
 & \le\left(1+\varepsilon\right)\left\Vert r^{(w)}_{i}-q^{(t)}_{i}\right\Vert _{2}\\
 & =\left(1+\varepsilon\right)\left\Vert l^{(t)}_{i}\right\Vert _{2}
\end{aligned}
\]
\end{proof}

Having proved Lemma~\ref{lem:block-norm-preserve}, we can now prove
Lemma~\ref{lem:greedy-almost-optimal}.
\begin{proof}
We know that $\Vert l^{(t)}_{i}\Vert_{2}$ is the inverse of the distance
of $A_{1:m,i}$ to the span of the other columns not yet chosen~\cite{doi:10.1016/j.apnum.2007.01.008}.
Lemma~\ref{lem:block-norm-preserve} guarantees that the column we
chose, which minimized $\Vert\widetilde{l}^{(t)}_{i}\Vert_{2},$ has
true residual norm $\Vert l^{(t)}_{i}\Vert_{2}$ within a factor of
$\frac{1+\varepsilon}{1-\varepsilon}$ from the minimal residual norm.
Since $\Vert l^{(t)}_{i}\Vert_{2}$ is the inverse of the distance
of $A_{1:m,i}$ to $\Span{A_{1:m,U_{t}\setminus\left\{ i\right\} }}$,
this proves Lemma~\ref{lem:greedy-almost-optimal}.
\end{proof}

The second property is a probabilistic version of another property
of V-BLAST, namely that it maximizes the magnitude of the smallest
diagonal element in the $R$ factor of $AP$ over all permutations:
\begin{lemma}
For every permutation $P$, denote \label{lem:almost-best-r-diagonal}
\[
R\left(P\right)=\min_{i}\left|R_{i,i}\right|\;\text{such that}\;AP=QR
\]
Let $P_{\text{R-V-BLAST}}$ denote a permutation matrix returned by
our algorithm. Let $0<\delta<\nicefrac{1}{2}$ and $0<\varepsilon<1$.
If we choose the projection dimension 
\[
b+p\ge C\left(\frac{b\log n+\log\frac{1}{\delta}}{\varepsilon^{2}}\right)
\]
for some absolute constant C, then with probability at least $1-\delta$,

\[
R\left(P_{\text{R-V-BLAST}}\right)\ge\frac{1-\varepsilon}{1+\varepsilon}\cdot\max_{P}R\left(P\right)
\]
\end{lemma}

\begin{proof}
First, we introduce some notation. The notation for $R(P)$ kept the
matrix $A$ implicit. For the purpose of the proof, it is useful to
make the $A$ matrix explicit, i.e., we denote
\[
R\left(A,P\right)=\min_{i}\left|R_{i,i}\right|\;\text{such that}\;AP=QR
\]
We also denote 
\[
R_{i}\left(A,P\right)=\left|R_{i,i}\right|\;\text{such that}\;AP=QR.
\]
Given a sequence of pivot selections $c_{1},\dots,c_{n}$, it defines
a unique permutation $P_{c}$ which puts column $c_{1}$ last, $c_{2}$
second last, and so on. Conversely, every permutation $P$ defines
a unique sequence of pivot selections $c_{1},\dots,c_{n}$ and the
corresponding sets $U_{0},\dots,U_{n}$. To keep the proof concise,
we will drop the subscript from $P_{c}$, and take it as understood
that $P$ corresponds to pivots $c_{1},\dots,c_{n}$ and sets $U_{0},\dots,U_{n}$,
while $P^{\prime}$ corresponds to pivots $c^{\prime}_{1},\dots,c^{\prime}_{n}$
and sets $U^{\prime}_{0},\dots,U^{\prime}_{n}$. Observe \cite{doi:10.1016/j.apnum.2007.01.008}:
\begin{equation}
R_{n-t}(A,P)=\left\Vert A_{1:m,c_{t+1}}-\Span{A_{1:m,U_{t}\setminus\left\{ c_{t+1}\right\} }}\right\Vert _{2}~.\label{eq:Rn-t}
\end{equation}
Another important property is monotonicity of $R(A,P)$. For every
subset $U$, define $P_{|U}$ to be the $|U|$-by-$|U|$ permutation
matrix obtained by deleting the columns of $P$ whose $1$ lies outside
the rows indexed by $U$, and then deleting those rows. The remaining
rows and columns retain their original order. So we have \cite{doi:10.1109/issse.1998.738086}
\[
R(A_{1:m,U},P_{|U})\geq R(A,P)~.
\]

We prove that any permutation $P'$ that upholds Eq.~(\ref{eq:lem1-property})
satisfies

\[
R\left(P^{\prime}\right)\ge\frac{1-\varepsilon}{1+\varepsilon}\cdot\max_{P}R\left(P\right)
\]
Since Lemma \ref{lem:greedy-almost-optimal} guarantees that Eq.~(\ref{eq:lem1-property})
holds with probability at least $1-\delta$ for $P_{\text{R-V-BLAST}}$
we see that if we choose projection dimension according to Lemma~\ref{lem:almost-best-r-diagonal}
preconditions, our claim will be proved.

We prove the claim by induction on $n$. When $n=1$ the claim trivially
holds (there is only one column to choose). Suppose this claim is
correct for all matrices with at most $n$ columns. Let $A$ be a
matrix with $n+1$ columns and $m$ rows. Let $P^{\prime}$ be a permutation
which upholds Eq.~(\ref{eq:lem1-property}). Let $P$ be any permutation,
and consider $t=0$ in Eq.~(\ref{eq:lem1-property}). Using Eq.~(\ref{eq:Rn-t}),
and since some column is placed last by $P$, Eq.~(\ref{eq:lem1-property})
implies that 
\[
R_{n+1}(A,P^{\prime})\geq\frac{1-\varepsilon}{1+\varepsilon}\cdot R_{n+1}(A,P)~.
\]
Since Eq.~(\ref{eq:lem1-property}) only depends on the sets $U_{t}$
and not on the order of the previous pivots, we have that $P^{\prime}_{|U^{\prime}_{1}}$
also upholds Eq.~(\ref{eq:lem1-property}) for $A_{1:m,U^{\prime}_{1}}$.
Thus, we have
\[
R(A_{1:m,U^{\prime}_{1}},P^{\prime}_{|U^{\prime}_{1}})\geq\frac{1-\varepsilon}{1+\varepsilon}\cdot R(A_{1:m,U^{\prime}_{1}},P_{|U^{\prime}_{1}})\geq\frac{1-\varepsilon}{1+\varepsilon}\cdot R(A,P)
\]
The first inequality is the inductive assumption, and the second is
monotonicity. Finally, 
\[
\begin{aligned}R(A,P^{\prime}) & =\min(R_{n+1}(A,P^{\prime}),R(A_{1:m,U^{\prime}_{1}},P^{\prime}_{|U^{\prime}_{1}}))\\
 & \geq\min\left(\frac{1-\varepsilon}{1+\varepsilon}\cdot R_{n+1}(A,P),\frac{1-\varepsilon}{1+\varepsilon}\cdot R(A,P)\right)\\
 & =\frac{1-\varepsilon}{1+\varepsilon}\cdot R(A,P)
\end{aligned}
\]
\end{proof}

\subsection{Analyses of Parallelism and Cache Misses}

We now analyze the algorithm from Section~\ref{sec:R-V-BLAST} in
terms of critical path and cache misses. We again ignore the computation
of $L=R^{-T}$ since it has insignificant span and cache misses relative
to the actual algorithm. 

The algorithm consists of $\lceil\frac{n}{b}\rceil$ phases that each
select the positions of up to $b$ columns in the ordering. Phase
$i$ operates on the remaining matrix, containing $n_{i}=n-(i-1)b$
rows and columns, reducing it to a sketch with dimensions $(b+p)\times n_{i}$
(possibly smaller in the last phase). Our performance analysis focuses
on the first $\lceil n/b\rceil/2$ phases, in which $n_{i}=\Theta(n)$.
Each of these phases consists of three subphases:
\begin{enumerate}
\item The algorithm generates a random $\left(b+p\right)\times n_{i}$ operator
and applies it to the remaining $n_{i}\times n_{i}$ matrix, to produce
the sketch. 
\item The algorithm runs $b$ steps of the Householder-based algorithm from
Section~\ref{sec:VBLAST-via-Householder} on the sketch, selecting
and ordering $b$ columns.
\item The algorithm aggregates the $b$ columns (in the remaining submatrix,
not in the sketch) to a block reflector, as in blocked (partitioned)
QR factorizations, and applies the block reflector to the remaining
submatrix.
\end{enumerate}
We again begin with an analysis of parallelism. The span of the first
subphase is $\Theta(\log n)$ whether we use a dense sketching operator
or one based on fast transforms. The span of the second subsubphase
is $\Theta(b\log n)$; this follows from our analysis of the Householder-based
algorithm. The span of the third subphase is also $O(b\log n)$. Given
that the number of phases is $\Theta(n/b)$, the total span is $\Theta(n\log n$),
asymptotically the same as the Householder-based and the downdating
algorithms.

However, the cache efficiency of this algorithm can be much better
than that of the Householder-based and the downdating algorithms.
We assume, as before, that $M\leq n^{2}/8$, and we also assume for
simplicity that $p=O(b)$ (setting $p=\Theta(b\log n)$ perturbs the
following estimates, but not enough to justify the added complexity).
Both the first and the third subphases access the entire $n_{i}\times n_{i}$
remaining submatrix, so over all the phases, they perform $\Omega((n/b)n^{2})=\Omega(n^{3}/b)$
cache misses. The third substep applies a block Householder reflection,
and this requires $\Theta(n^{2}_{i}b/\smash{\sqrt{M}})$ cache misses
per phase, or $\Theta(n^{3}/\smash{\sqrt{M}})$ over the entire algorithm.
If the sketching operator is dense, multiplying it by the panel of
the remaining submatrix also incurs $\Theta(n^{2}_{i}b/\smash{\sqrt{M}})$
cache misses per phase. If it is implemented using fast transform,
the number of cache misses is generally smaller, so we ignore this
term. Either way, the cost of the first subphase dominates.

We now analyze the number of cache misses generated by applying the
Householder-based algorithm to each sketch. If $(b+p)n\leq cM$ for
some appropriate $c<1$, so the sketch fits within a fraction of the
cache (a typical choice of parameters for panel-based algorithms),
then the cache misses required to read the sketch, $O((b+p)n)$, dominate,
for a total cache miss count in these subphases over the entire algorithm
of $O(n^{2})$, clearly a low-order term. If the sketch does not fit
within the cache, we may need to read the (updated) sketch before
we select each column, increasing the upper bound on the cache-miss
count by a factor of $b$ to $O(bn^{2}$).

To summarize, if the sketches fit within the cache, the total cache
miss count is 
\[
\Theta\left(\frac{n^{3}}{b}+\frac{n^{3}}{\sqrt{M}}+n^{2}\right)
\]
and 
\[
\Theta\left(\frac{n^{3}}{b}+\frac{n^{3}}{\sqrt{M}}+bn^{2}\right)
\]
otherwise.

\section{Experimental Results}\label{sec:Experimental-Results}

Experiments were performed on a laptop with a 4-core / 8-thread 2.60
GHz Intel Core i5-1145G7 CPU and 32~GB of RAM. We implemented all
the algorithms in C with calls to the BLAS~\cite{doi:10.1145/77626.79170,doi:10.1145/42288.42291}
and LAPACK~\cite{doi:10.1137/1.9780898719604} where appropriate.
We linked the code with Intel's Math Kernel Library (MKL), which contained
highly optimized implementations of many BLAS and LAPACK routines.
The version of MKL that we used is part of Intel's OneAPI version
2026.1. Appendix~\ref{sec:appendix-code} presents additional details
about our implementations. We performed the experiments presented
below using Python code that invoked the C routines. We ran the code
on one core using one thread, in order to make the running times and
cache-miss counts easy to interpret.

We conducted two groups of experiments. The first group, described
in the next section, aims to characterize the performance of our new
algorithms relative to the performance of existing V-BLAST algorithms
(and relative to relevant baselines, like a simple QR factorization).
The second group analyzes the effectiveness of exact and approximate
V-BLAST orderings in the context of a full integer least-squares solver.

\subsection{The Efficiency of Our Algorithms}

We compare the performance of several exact and approximate V-BLAST
algorithms:
\begin{itemize}
\item Our new Householder-based algorithm from Section~\ref{sec:VBLAST-via-Householder},
which computes the exact V-BLAST ordering, 
\item our new random-projections algorithm from Section~\ref{sec:R-V-BLAST},
which produces an approximation to the V-BLAST ordering,
\item the exact V-BLAST algorithm by Chang and Paige~\cite{doi:10.1016/j.apnum.2007.01.008},
\item and an exact downdating algorithm based on the work of Benesty, Huang,
and Chen~\cite{doi:10.1109/tsp.2003.812897}.
\end{itemize}
We implemented all the algorithms ourselves, since we did not find
existing high-performance implementations of them. For the random
sampling algorithm, we chose block size $b=16$ and oversampling parameter
$p=6$.

\begin{figure}[tb]
\centering{}\includegraphics[width=0.6\textwidth]{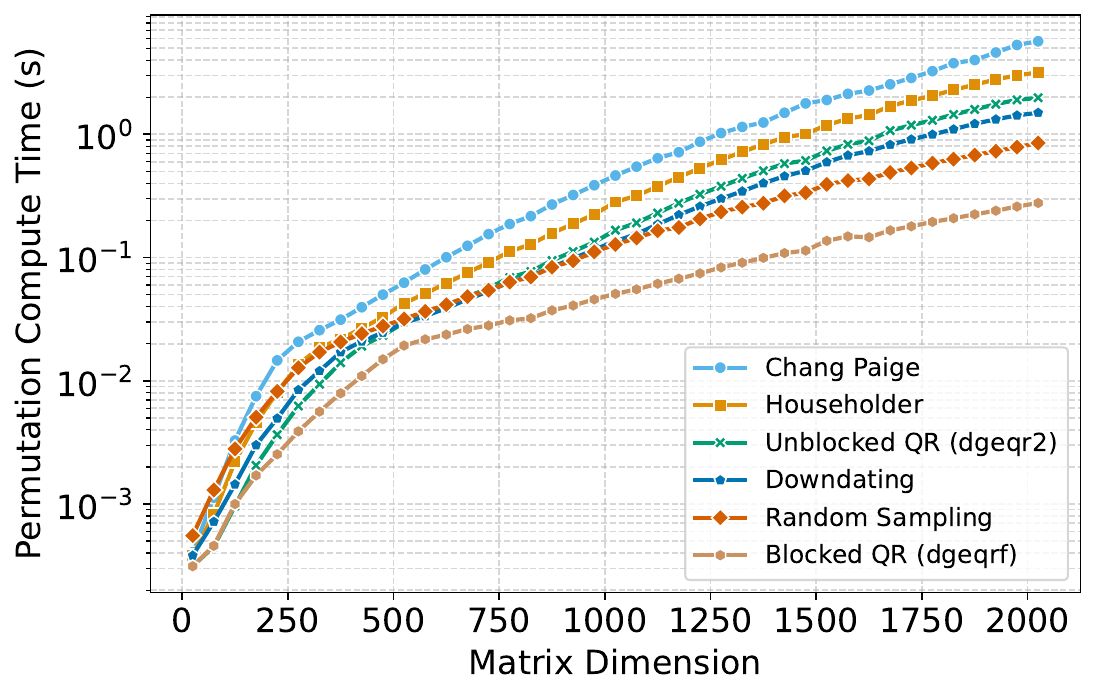}\caption{Average running times of exact and approximate V-BLAST algorithms,
as well as of two QR factorizations included for calibration. Each
marker represents an average of 100 runs on different Gaussian random
matrices.}\label{fig:perm-time-vs-size}
\end{figure}

The running times of these algorithms are presented in Figure~\ref{fig:perm-time-vs-size}.
The figure also includes the running times of LAPACK's \texttt{dgeqr2}
(BLAS~2-based QR factorization) and \texttt{dgeqr}f (BLAS~3-based
QR) routines for calibration. Our Householder V-BLAST algorithm is
structurally similar to \texttt{dgeqr2} in the sense that it computes
and applies $n-1$ Householder reflections to a matrix. Our repeated
random sampling algorithm is structurally similar to \texttt{dgeqrf}
in the sense that both apply a series of block reflections to a matrix. 

The running times presented in Figure~\ref{fig:perm-time-vs-size}
are averages of 100 runs on different random matrices with Gaussian
i.i.d. entries. In general, the variance of the running times is fairly
small; we present data to support this observation in Figure~\ref{fig:variance-vs-dim}
of Appendix~\ref{sec:appendix-graphs}.

The graphs show that our Householder algorithm is about $50\%$ slower
than LAPACK's \texttt{dgeqr2 }(which is structurally similar but does
not need to search for columns with extreme norms). The algorithm
of Chang and Paige is much slower, taking about $125\%$ more time
than \texttt{dgeqr2} (worse than double the running times).

The downdating algorithm is faster than all three: it is faster by
$25\%$ than \texttt{dgeqr2}. It takes roughly half the running time
of our Householder algorithm, but it is not as numerically stable.

Our new random-sampling algorithm is faster than all the exact V-BLAST
algorithms on matrices with dimension above about 750, and is faster
than the Householder algorithm above dimension about 500, but on smaller
matrices it is not the fastest. 

\begin{figure}[tb]
\centering{}\includegraphics[width=0.6\textwidth]{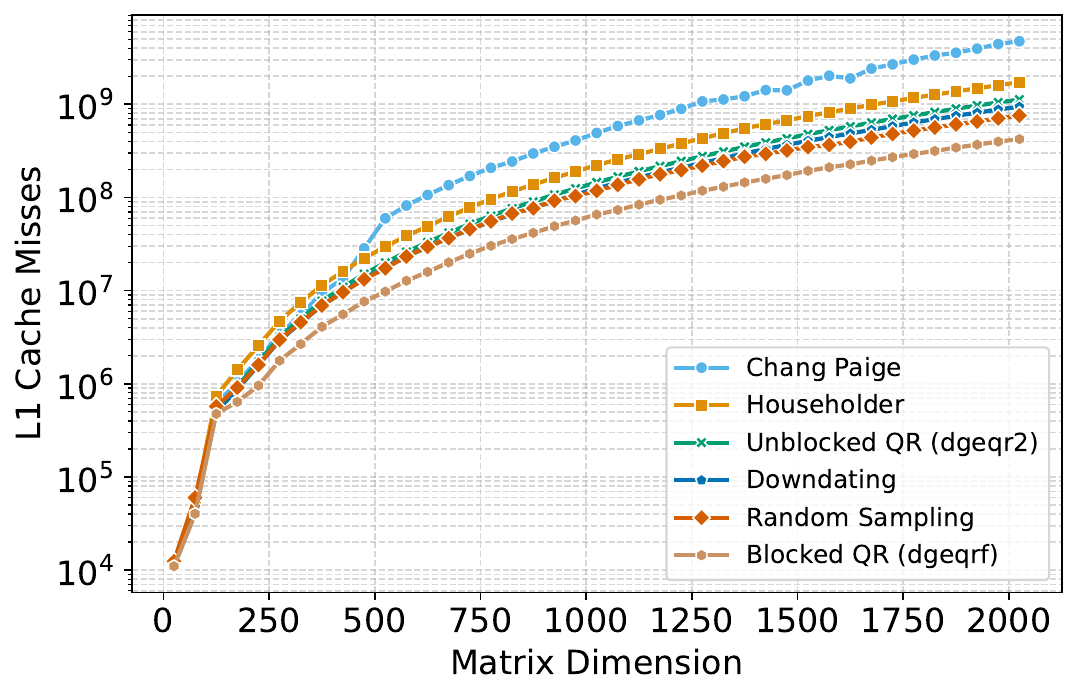}\caption{Average numbers of L1 cache misses for exact and approximate V-BLAST
algorithms.}\label{fig:l1-cache-misses-vs-size}
\end{figure}

\begin{figure}[tb]
\centering{}\includegraphics[width=0.6\textwidth]{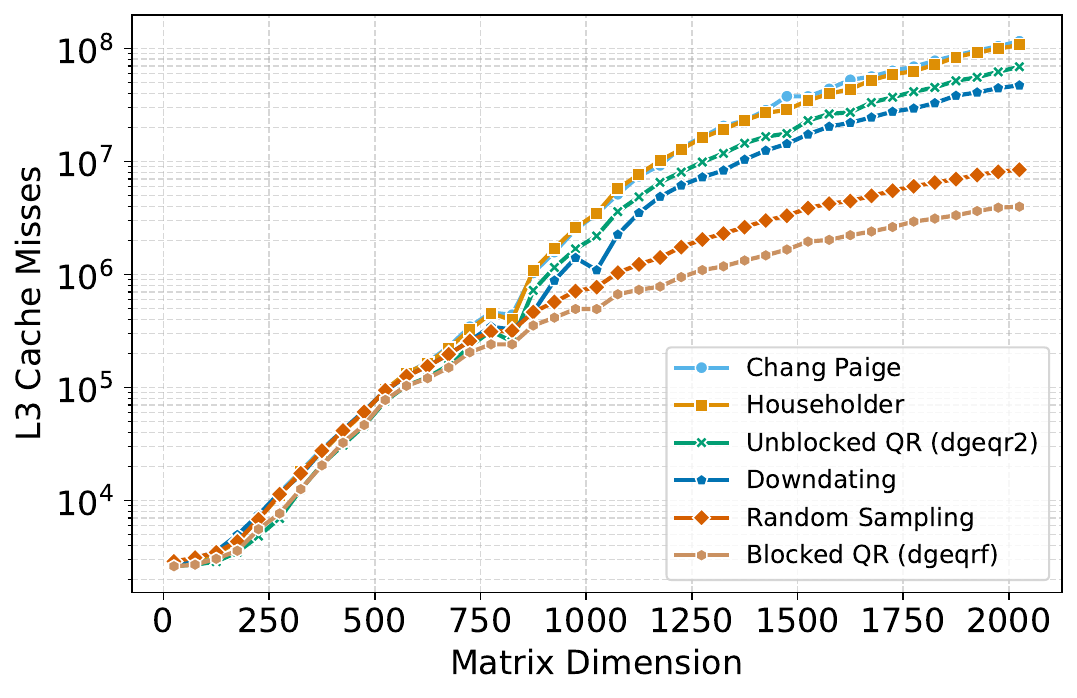}\caption{Average numbers of L3 cache misses for exact and approximate V-BLAST
algorithms.}\label{fig:l3-cache-misses-vs-size}
\end{figure}
Figures~\ref{fig:l1-cache-misses-vs-size} and~\ref{fig:l3-cache-misses-vs-size}
present graphs that display the average number of cache misses that
the algorithms generate. Cache-miss counts were read from the CPU's
hardware performance counters by dynamically attaching the Linux \texttt{perf}
utility to the process running the experiment. The dynamic attachment
ensures that the counts only reflect cache misses incurred during
the execution of the algorithm and not in the setup or teardown phases.
The graphs present miss counts from the first- and last-level caches
(L1 and L3), which are relatively straightforward to count with \texttt{perf}
(L2 misses are more difficult to count). 

The results show that as claimed, our random sampling algorithm incurs
far fewer cache misses than the exact V-BLAST algorithms. This is
particularly pronounced in the L3 counts, where the cache-miss curve
of the randomized algorithm tracks that of the Level-3 BLAS \texttt{dgeqrf},
whereas the exact algorithms all behave qualitatively differently
starting at about dimension 800, where their behavior starts to track
the Level-2 BLAS \texttt{dgeqr2}.

The L1 graphs mostly emphasize the poor cache performance of the algorithm
of Chang and Paige, which degrades significantly at about dimension
500.

\subsection{Evaluation of the Quality of Approximate V-BLAST Permutations}

Next, we present experiments that quantify the effects of approximating
the V-BLAST permutation on an overall integer least-squares solver.
We evaluate two approximations: our random-sampling algorithm and
SQRD, a simpler column-pivoting QR algorithm~\cite{doi:10.1049/el:20010899}.

We evaluate all three permutations in a solver that produces an exact
minimizer (up to the effect of floating-point rounding errors) to
the problem

\[
\hat{x}=\arg\min_{x\in\mathbb{Z}^{n}}\left\Vert Ax-b\right\Vert _{2}
\]
where $A$ is a random matrix with standard Gaussian i.i.d. entries
and where $b=A\cdot y+e$ with $y$ being an integer vector generated
by rounding a standard Gaussian i.i.d. random vector and $e$ a Gaussian
noise vector with i.i.d. entries with a prescribed standard deviation.
The choice of distribution for the coefficient matrix $A$ implies
that it has full column rank with high probability. The choice of
the standard deviation for $e$ has a strong influence on how hard
the least-squares problem is, as we shall see below.

The solver finds $\hat{x}$ by computing the QR factorization $A=QR$,
computing an exact or approximate V-BLAST column permutation $P$
of $R$, computing the QR factorization $AP=Q_{1}R_{1}$, and then
invoking the Schnorr-Euchner (SE) search to solve 
\begin{eqnarray*}
\hat{x} & = & \arg\min_{x\in\mathbb{Z}^{n}}\left\Vert Ax-b\right\Vert _{2}\\
 & = & \arg\min_{x\in\mathbb{Z}^{n}}\left\Vert AP\left(P^{T}x\right)-b\right\Vert _{2}\\
 & = & \arg\min_{x\in\mathbb{Z}^{n}}\left\Vert Q_{1}R_{1}\left(P^{T}x\right)-b\right\Vert _{2}\\
 & = & \arg\min_{x\in\mathbb{Z}^{n}}\left\Vert R_{1}\left(P^{T}x\right)-Q^{T}_{1}b\right\Vert _{2}\;.
\end{eqnarray*}
The running time of the first QR factorization is essentially independent
of all the random choices, and the running time of determining $P$
is also largely independent of them. But the running time of the SE
search depends strongly on both the standard deviation of $e$ and
on the quality of $P$. 

\begin{figure}[tb]
\centering{}\includegraphics[width=0.6\textwidth]{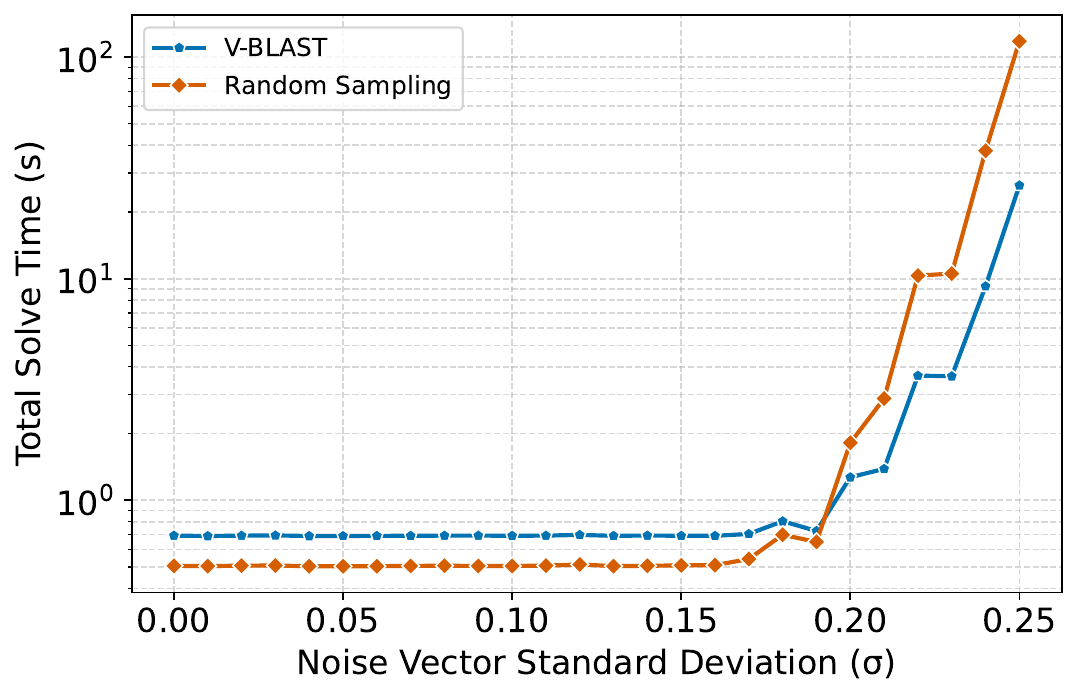}\caption{Average total solving time (both the permutation computation and
the SE algorithm) for a $1500\times1500$ matrix as a function of
the standard deviation of elements of $e$, the additive error in
the right--hand side $b$. Each marker represents the average of
$100$ matrices.}\label{fig:total-time-vs-noise}
\end{figure}

Figure~\ref{fig:total-time-vs-noise} explores the effect of the
standard deviation $\sigma$ of elements of $e$ on the total solution
time. Up to about $\sigma=0.15$, the running times with both exact
V-BLAST and our randomized algorithm are independent of $\sigma$.
In this regime, the SE search is very fast and the QR factorizations
and the construction of the permutation dominate the running time.
In this regime, the faster but approximate randomized algorithm leads
to faster solution times. As $\sigma$ grows beyond 0.15, the search
phase becomes more and more expensive, the running times rise quickly,
and the solver that uses an exact V-BLAST permutation becomes faster.

Therefore, from here on we use a fixed standard deviation $\sigma=0.15$
for elements of $e$. This is close to the largest possible noise
for which the search is still inexpensive at dimension $1500$. 

\begin{figure}
\includegraphics[width=0.45\textwidth]{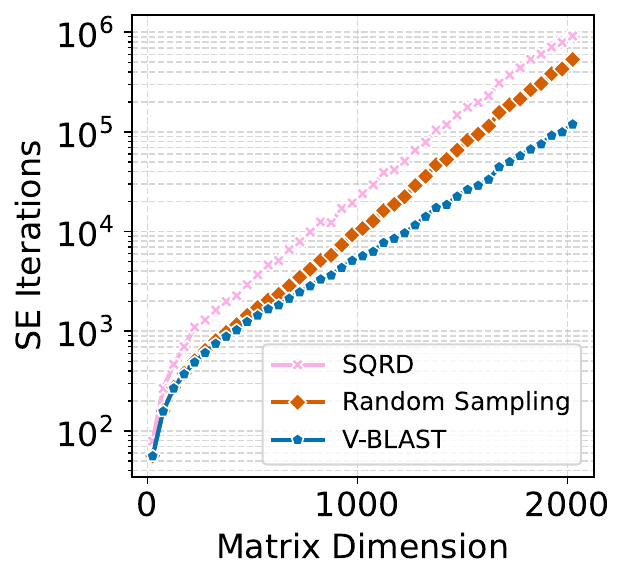}\hspace*{\fill}\includegraphics[width=0.45\textwidth]{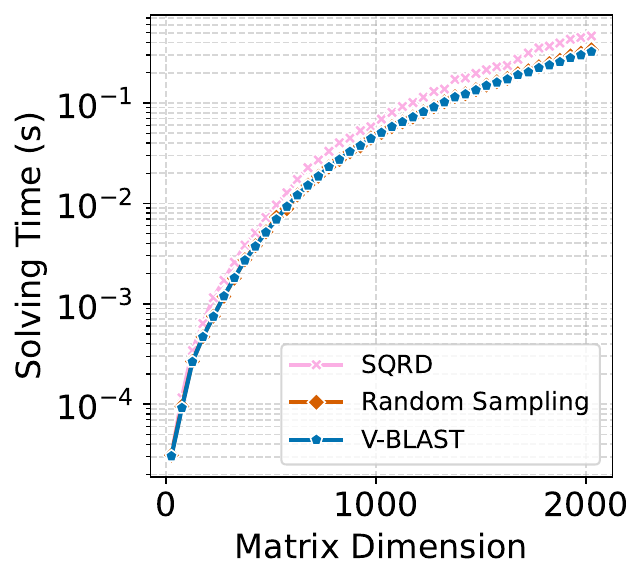}\caption{SE iteration counts and total running times of the solver, excluding
permutation computation, as a function of problem dimension. Each
marker represents a geometric mean of $1000$ experiments. The standard
deviation of elements of the right-hand side error is $\sigma=0.15$.}\label{fig:se-metrics-vs-size}
\end{figure}

Figure~\ref{fig:se-metrics-vs-size} shows how the SE iteration counts
and the solver running times, excluding permutation computation, scale
with matrix dimension for the three types of permutations (V-BLAST,
SQRD, and our randomized algorithm), all for noise vectors whose elements
have standard deviation $\sigma=0.15$. We limited the number of SE
iterations to $25n^{2}$, to keep the running times reasonable. Although
SE may theoretically need an exponential number of iterations, the
relatively small standard deviation of the noise allows SE to converge
to a minimizer within this bound in most runs. With the V-BLAST and
our randomized ordering it converged in at least 98\% of the runs,
and with SQRD in at least 90\%. The SE iteration counts are explored
in more detail later in this section with even more details in Appendix~\ref{sec:appendix-graphs}. 

The data indicate that our randomized algorithm (again with block
size $b=16$ and oversampling parameter $p=6$) produces permutations
that are less effective than exact V-BLAST but more effective than
SQRD. At dimension $2000$ SQRD leads to about 10 times more iterations
than V-BLAST; this factor grows with matrix dimension. 

The running times differ much less. The running times of the solver
with exact V-BLAST and with the permutation produced by the randomized
algorithm are similar. Those that use the SQRD permutation are slower. 

\begin{figure}
\includegraphics[width=0.45\textwidth]{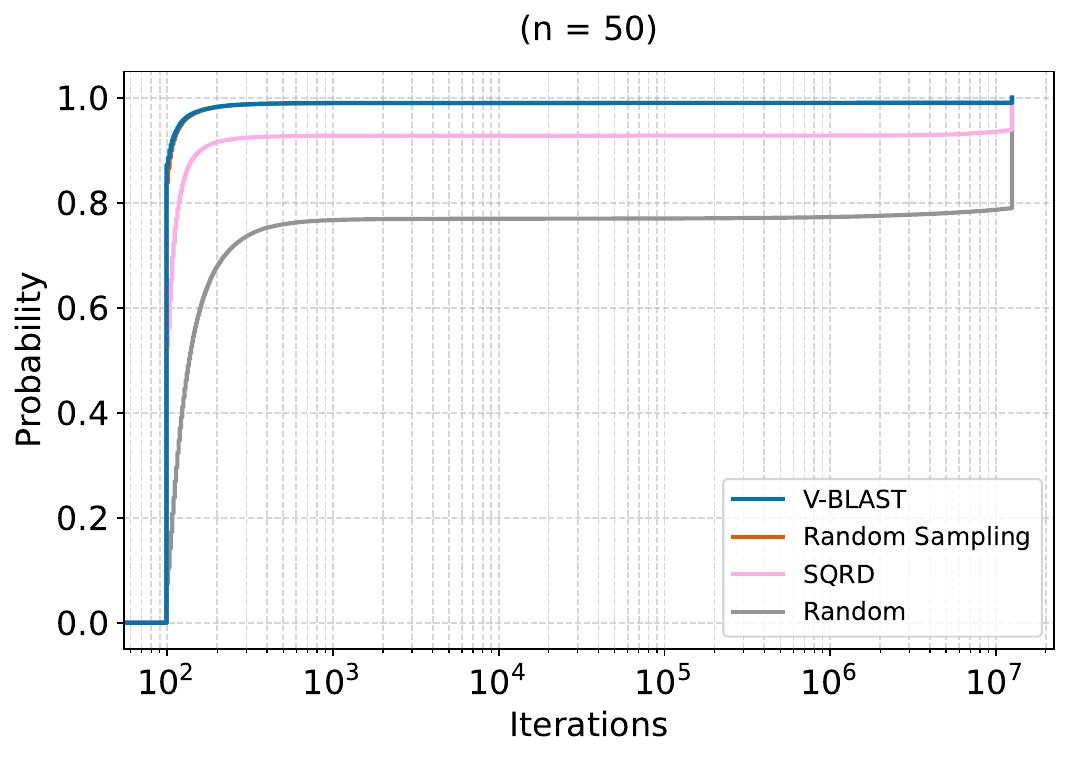}\hspace*{\fill}\includegraphics[width=0.45\textwidth]{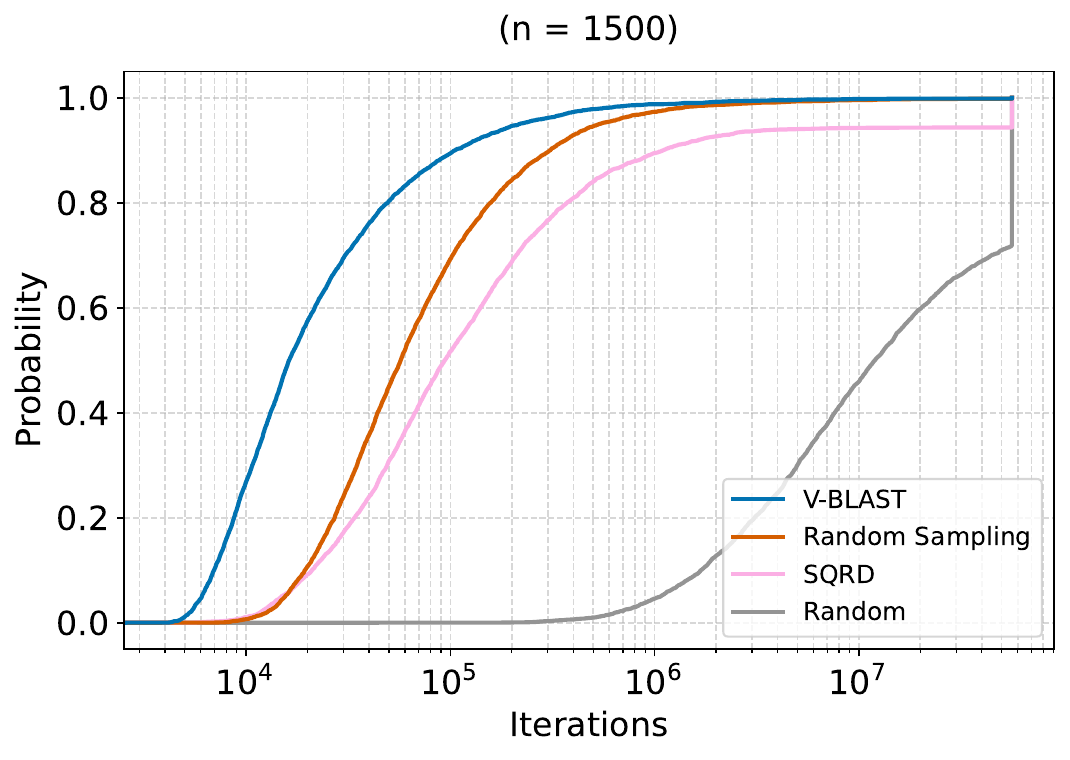}\caption{ECDF of the iterations needed by SE for different permutation methods
for matrices of dimensions $50$ (left) and $1500$ (right). The number
of iterations was limited to $1.25\cdot10^{7}$ and $5.625\cdot10^{7}$,
respectively.}\label{fig:iterations-ecdf-no-subs}
\end{figure}

Representing the distributions of iteration counts using only (geometric)
means in Figure~\ref{fig:se-metrics-vs-size} hides some important
details. Figure~\ref{fig:iterations-ecdf-no-subs} presents the empirical
cumulative distributions for two problem dimensions, $50$ and $1500$.
The data is based on 100,000 and 5,000 experiments per dimension,
respectively, and on noise whose elements have standard deviation
$\sigma=0.15$. 

The data show that better ordering methods are better in two distinct
ways. First, under all the ordering methods, some fraction of problem
instances exhibits very slow convergence of the SE search (regions
in which the graphs appear to be flat). For $n=50$, most of the instances
that did not converge in 1,000 iterations did not converge in $10^{7}$
iterations. Without the imposed iteration limit, all runs would eventually
converge (SE terminates), but some problem instances remain hard even
with the given ordering. The fraction of these stubborn instances
is smallest for V-BLAST and random sampling, is much larger for SQRD,
and is largest for a random column ordering. 

Second, for problem instances that are not stubborn, V-BLAST usually
requires fewer iterations than the randomized ordering, which in turn
requires fewer iterations than SQRD. This is evident from the region
of iteration counts in which the graphs are rapidly increasing.

\section{Conclusions}\label{sec:Conclusions}

More than 60 years after the publication of Wilkinson's \emph{Rounding
Errors in Algebraic Processes~\cite{doi:10.1137/1.9781611977523}}
and almost 25 years after the publication of the 2nd edition of Higham's
\emph{Accuracy and Stability of Numerical Algorithms}~\cite{doi:10.1137/1.9780898718027},
the state of the art in algorithms for dense matrix computations is
very well established. For most problems, we have cubic-complexity
backward-stable algorithms with a critical path (span) of $O(n\log n)$,
where $n$ is the dimension, with good reuse of data in caches and
local memories, and in many cases also cache-oblivious and communication-avoiding
algorithms, which bring these data-movement metrics to extreme asymptotic
efficiency. Many of the most stable algorithms are based on orthogonal
transformations. There are also sub-cubic algorithms for most problems,
but they are often not as stable (and hard to make fast in practice).

Algorithms for computing the V-BLAST permutation, which belongs to
this class of problems, lagged behind. While a cubic algorithm was
discovered in 2000~\cite{doi:10.1109/icassp.2000.859065} and again
independently in 2007~\cite{doi:10.1016/j.apnum.2007.01.008}, parallelism,
data reuse, and numerical stability have not been analyzed for any
of them.

Our article rectifies this situation. We have analyzed the parallelism
and data reuse in existing algorithms, shown that they are lacking,
proposed two new algorithms with improved combinations of parallelism,
stability, and data reuse, and we have shown both theoretically and
experimentally that they are effective.

One of the new algorithms that we proposed is exact, and our results
indicate that it is superior to existing exact V-BLAST algorithms.
The other is approximate, but the approximation is highly effective
and its performance is even better. We, therefore, recommend that
one of them is used in lieu of existing algorithms whenever possible.

Both of our new algorithms use algorithmic strategies borrowed from
strategies that have been used in algorithms that compute rank-revealing
QR factorizations. The two problems are closely related (but not identical),
which allowed us to adapt these strategies to V-BLAST. The resulting
algorithms are structurally similar to the corresponding algorithms
for computing rank-revealing QR factorizations. We, therefore, believe
that the performance of our new algorithms on parallel hardware (multicore
CPUs, GPUs, etc.) would track the performance of the corresponding
algorithms for rank-revealing QR.

Finally, there is a class of high-performance algorithms for rank-revealing
QR, namely algorithms based on tournaments~\cite{doi:10.1137/13092157X},
for which we have not found an equivalent for V-BLAST. Is there a
way to compute an approximate V-BLAST permutation efficiently using
a tournament-based algorithm?

\appendix

\section{Implementation Details and Documentation of the Code}\label{sec:appendix-code}

We implemented all the algorithms initially in Python without any
use of AI. These implementations emphasize clarity and we used them
to verify the correctness of the algorithms. We then used Gemini to
improve the performance of the Python code and to convert the algorithms
to C. We tested all the high-performance implementations by comparing
their outputs to those of the plain-vanilla hand-written versions. 

Our code can be found here \url{https://github.com/TomerPelleg/V-BLAST}.
The main components of our code are:
\begin{itemize}
\item \texttt{columns\_ordering\_algorithms} directory: this directory contains
C and Python implementations for the various V-BLAST algorithms, for
our repeated sampling algorithm, for SQRD, and for wrappers for \texttt{dgeqrf}
and \texttt{dgeqr2}. It also contains the CMake files for the project,
and a Python permutation class used to wrap permutation results. The
\texttt{exact\_vblast/main.py} contains tests that check whether different
functions produce the same permutation on matrices it generates. This
proved particularly useful for checking that we understood all the
V-BLAST algorithms, since we could not find implementations for them.
\item \texttt{schnorr\_euchner} directory: this directory contains C and
Python implementations of the Schnorr-Euchner algorithm.
\item \texttt{experiments} directory: this directory contains the Python
code for running the experiments. It has code to generate matrices
and vectors and to run an experiment from end to end. It also has
a configuration class for the experiments which tells the experiment
runner which parameters to use, and the code used to attach \texttt{perf}
to each process. In the \texttt{visuals} sub-directory there is code
to generate the plots shown in this article. The \texttt{configs}
sub-directory has the configurations used for these plots.
\item \texttt{tests} directory: development tests for the project, based
on \texttt{pytest} testing framework.
\end{itemize}
We also note the main BLAS and LAPACK routines used by our implementations.
To compute $L=R^{-T}$, which was a common preprocessing step across
various algorithms, we used \texttt{dgeqrf} to compute the $R$ factor
in $A=QR$, and \texttt{dtrtri} to invert the triangular matrix and
compute $R^{-1}$. We used \texttt{dswap} to interchange columns.
\begin{itemize}
\item Chang and Paige V-BLAST algorithm: We used the above method to compute
$L=R^{-T}$. We used \texttt{drot} to apply Givens rotations.
\item Downdating V-BLAST algorithm: We used \texttt{dsyrk} to compute $A^{*}A$.
We used \texttt{dpotrf} and \texttt{dpotri} to compute the Cholesky
factorization of $A^{*}A$ and invert it, to get $M=\left(A^{*}A\right)^{-1}$.
We used \texttt{dger} for the rank-1 downdating of $M$. 
\item Householder V-BLAST algorithm (introduced in this article): We used
the above method to compute $L=R^{-T}$. We used \texttt{dgemv} and
\texttt{dger} to compute the Householder reflectors.
\item Repeated Random Sampling algorithm (introduced in this article): We
used the above method to compute $L=R^{-T}$. We used \texttt{dgemm}
to multiply $L$ with the random matrix $G$. Like the Householder
algorithm, we used \texttt{dgemv} and \texttt{dger} to apply the Householder
reflectors in the projected space. We used \texttt{dnrm2} to compute
the norms of the columns in the projected space. To build the block
reflectors, we used \texttt{dgemv}, \texttt{dger}, \texttt{dscal},
\texttt{dcopy} and \texttt{dnrm2}, and used \texttt{dgemm} to apply
them. This was done in a manner similar to the implementation of \texttt{dgeqrf}.
\item SQRD algorithm: We used \texttt{dgemv} and \texttt{dger} to compute
the Householder reflectors needed after finding the column with the
minimal norm.
\item Schnorr-Euchner: We used \texttt{dgeqrf} and \texttt{dormqr} to transform
the original least-squares problem with a general matrix $A$ to a
least-squares problem with a triangular matrix, as required by the
Schnorr-Euchner algorithm.
\end{itemize}
\textbf{AI usage description:} As stated above, the high-performance
implementations were generated by Gemini, using clear hand-written
Python code as a specification of the algorithms. The code generating
the graphs was also written with some use of AI.

\section{Additional Auxiliary Results}\label{app:aux}

In this appendix we prove some additional technical lemmas used in
proving the results of Section~(\ref{sec:theoretical}).

Recall that $S_{t}=\left[n\right]\setminus U_{t}$ denotes the set
of indices of the first $t$ columns selected by the algorithm. Recall
that $l^{(t)}_{i}$ is the orthogonal projection of $l_{i}$ onto
the orthogonal complement of $\Span{l_{j}|j\in S_{t}}$. Let $H$
be a unitary matrix that aggregates all the $n$ reflectors. It means
$H\left(LP\right)$ is some lower triangular matrix $T$. Let $c_{t}$
be the $t$ chosen column (meaning $\{c_{t}\}=S_{t}\setminus S_{t-1}).$
Denote by $h_{1},\dots,h_{n}$ be the rows of $H$ transposed into
a column vector (recall that $H$ is applied to $L$ which is $n$-by-$n$).
Since $H$ is unitary, we have $LP=H^{T}T$, or
\[
\left(LP\right)_{1:n,i}=\sum^{n}_{j=1}T_{j,i}\cdot h_{j}
\]

\begin{lemma}
\label{lem:gram-schmidt}
\[
h_{n-t+1}\parallel l^{(t-1)}_{c_{t}}
\]
\end{lemma}

\begin{proof}
Note that the definition of $c_{t}$, $l_{c_{t}}=(LP)_{1:n,n-t+1}$We
prove by induction on $t$. For the base case, first note that the
topmost $n-1$ elements in the rightmost column of $H\left(LP\right)$
are zero, so the bottom most entry, at position $n,n$, has absolute
value which equal the norm of $n$ column in $H\left(LP\right)$.
However, $H$ is unitary, so that norm is equal to the norm of column
$n$ in $LP$. Further, that entry is equal to to the dot product
of the $n$-th row of $H$ (i.e., $h_{n})$ with the $n$-th column
of $LP$, so $|h^{T}_{n}\cdot\left(LP\right)_{1:n,n}|$ is equal to
the norm of $\left(LP\right)_{1:n,n}$. Now, since $h_{n}$ has unit
norm (since it is a column of a unitary matrix), we conclude that
$h_{n}\parallel l^{(0)}_{c_{1}}$.

Since $LP=H^{T}T$ and the fact that $T$ is lower triangular
\[
(LP)_{1:n,n-t+1}=\sum^{n}_{j=n-t+1}T_{j,n-t+1}\cdot h_{j}
\]
where used the fact the top entries in that column of $T$ are $0$.From
induction $\sum^{n}_{j=n-t+2}T_{j,n-t+1}\cdot h_{j}$ is in $\Span{l_{j}|j\in S_{t-1}}$.
Since $h_{n-t+1}$ is orthogonal to all $h_{n-t+2},\dots,h_{n}$,
it means that $T_{n-t+1,n-t+1}\cdot h_{n-t+1}$is the orthogonal projection
of $(LP)_{1:n,n-t+1}$ into the orthogonal compliment of $\Span{l_{j}|j\in S_{t-1}}$,
which is by definition $l^{(t-1)}_{c_{t}}$.
\end{proof}

\begin{corollary}
\label{cor:span}$l^{(t)}_{i}\in\Span{h_{1},\dots,h_{n-t}}$
\end{corollary}

Let $w$ be a multiple of $b$. Recall that the algorithm aggregates
Householder transformation and applies them at such steps, and that
we denoted the by $H_{w}$ the aggregated transformation after selecting
$w$ columns. It is easy to see that bottom $w$ rows of $H$ are
the bottom $w$ rows of $H_{w}$. Thus, we can write

\[
H_{w}=\begin{bmatrix}\tilde{H}_{w}\\
H_{n-w+1:n,1:n}
\end{bmatrix}
\]
and by definition
\[
r^{(w)}_{i}=\tilde{H}_{w}l_{i}
\]

\begin{lemmastar}
$\lVert r^{(w)}_{i}\rVert_{2}=\Vert l^{(w)}_{i}\Vert{}_{2}$.
\end{lemmastar}

\begin{proof}
Let us write $l_{i}=\sum^{n}_{j=1}\alpha_{ij}h_{j}$ (i.e., expand
$l_{i}$ in the basis $h_{1},\dots,h_{n})$. Due to Corollary~\ref{cor:span}
and the fact that $h_{1},\dots,h_{n}$ is an orthonormal basis, we
see that $l^{(w)}_{i}=\sum^{n-w}_{j=1}\alpha_{ij}h_{j}.$We see that
\[
\lVert l^{(w)}_{i}\rVert^{2}_{2}=\sum^{n-w}_{j=1}|\alpha_{ij}|^{2}=\sum^{n}_{j=1}|\alpha_{ij}|^{2}-\sum^{n}_{j=n-w+1}|\alpha_{ij}|^{2}=\lVert l_{i}\|^{2}_{2}-\sum^{n}_{j=n-w+1}|\alpha_{ij}|^{2}
\]
On the other hand, 
\[
H_{w}l_{i}=\begin{bmatrix}r^{(w)}_{i}\\
\left(H_{n-w+1:n,1:n}\right)l_{i}
\end{bmatrix}=\begin{bmatrix}r^{(w)}_{i}\\
\alpha_{i,n-w+1}\\
\vdots\\
\alpha_{i,n}
\end{bmatrix}
\]
Since $H_{w}$ is unitary, 
\[
\lVert l_{i}\rVert^{2}_{2}=\lVert H_{w}l_{i}\rVert^{2}_{2}=\lVert r^{(w)}_{i}\rVert^{2}_{2}+\sum^{n}_{j=n-w+1}|\alpha_{ij}|^{2}
\]
The claim follows.
\end{proof}

\begin{lemmastar}
Denote by $q^{(t)}_{i}$ the orthogonal projection of $r^{(w)}_{i}$
into $\Span{r^{(w)}_{j}|j\in V_{t}}$. Since $r^{(w)}_{i}=(H_{w}l_{i})_{1:n-w}$
and $H_{w}$ is orthogonal, we have $\ensuremath{\left\Vert l^{(t)}_{i}\right\Vert _{2}=\left\Vert r^{(w)}_{i}-q^{(t)}_{i}\right\Vert _{2}}$.
\end{lemmastar}

\begin{proof}
Let $W_{t}$ be the product of the first $t-w$ Householder reflectors
used to process the current block in $L$, in the same order used
to form $H$. Thus, $W_{t}$ has size $(n-w)\times(n-w)$. Define
$\tilde{H}_{t}$ to be the first $n-t$ rows of $W_{t}\tilde{H}_{w}$.
Later reflectors leave the remaining rows unchanged, so we have
\[
W_{t}\tilde{H}_{w}=\begin{bmatrix}\tilde{H}_{t}\\
H_{n-t+1:n-w,1:n}
\end{bmatrix}
\]

\[
W_{t}r^{(w)}_{i}=W_{t}(\tilde{H}_{w}l_{i})=\begin{bmatrix}\tilde{H}_{t}l_{i}\\
H_{n-t+1:n-w,1:n}l_{i}
\end{bmatrix}=\begin{bmatrix}\tilde{H}_{t}l_{i}\\
\alpha_{i,n-t+1}\\
\vdots\\
\alpha_{i,n-w}
\end{bmatrix}
\]
where $\alpha_{i,n-t+1},\dots,\alpha_{i,n-w}$ are from the expansion
$l_{i}=\sum^{n}_{j=1}\alpha_{ij}h_{j}$ (see previous proof).So we
just need to show that

\[
q^{(t)}_{i}=W^{T}_{t}\begin{bmatrix}0\\
\alpha_{i,n-t+1}\\
\vdots\\
\alpha_{i,n-w}
\end{bmatrix}
\]
This follows because
\[
W_{t}\Span{r^{(w)}_{j}|j\in V_{t}}=\{0\}^{n-t}\times\mathbb{R}^{t-w}.
\]
Thus the displayed expression for $q^{(t)}_{i}$ is the orthogonal
projection onto $\Span{r^{(w)}_{j}|j\in V_{t}}$. By orthogonality
and Corollary~\ref{cor:span},
\[
\left\Vert r^{(w)}_{i}-q^{(t)}_{i}\right\Vert _{2}=\left\Vert \tilde{H}_{t}l_{i}\right\Vert _{2}=\left\Vert l^{(t)}_{i}\right\Vert _{2}.
\]
\end{proof}

\FloatBarrier

\section{Additional Experimental Results}\label{sec:appendix-graphs}

This section has some graphs that were not included in the main paper.

\begin{figure}[tbph]
\centering{}\includegraphics[width=0.6\textwidth]{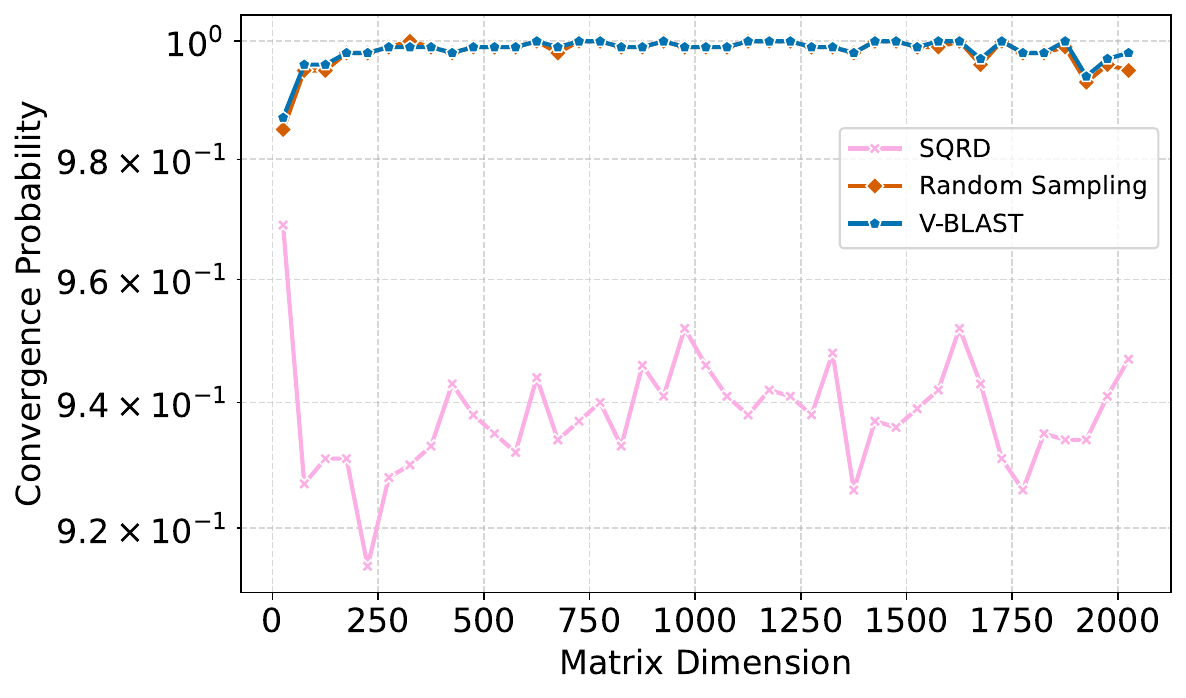}\caption{The percentage of experiments whose SE converged before reaching
the maximal number of iterations $25\cdot n^{2}$}\label{fig:convergence-prob-vs-size}
\end{figure}

\begin{figure}[tbph]
\centering{}\includegraphics[width=0.6\textwidth]{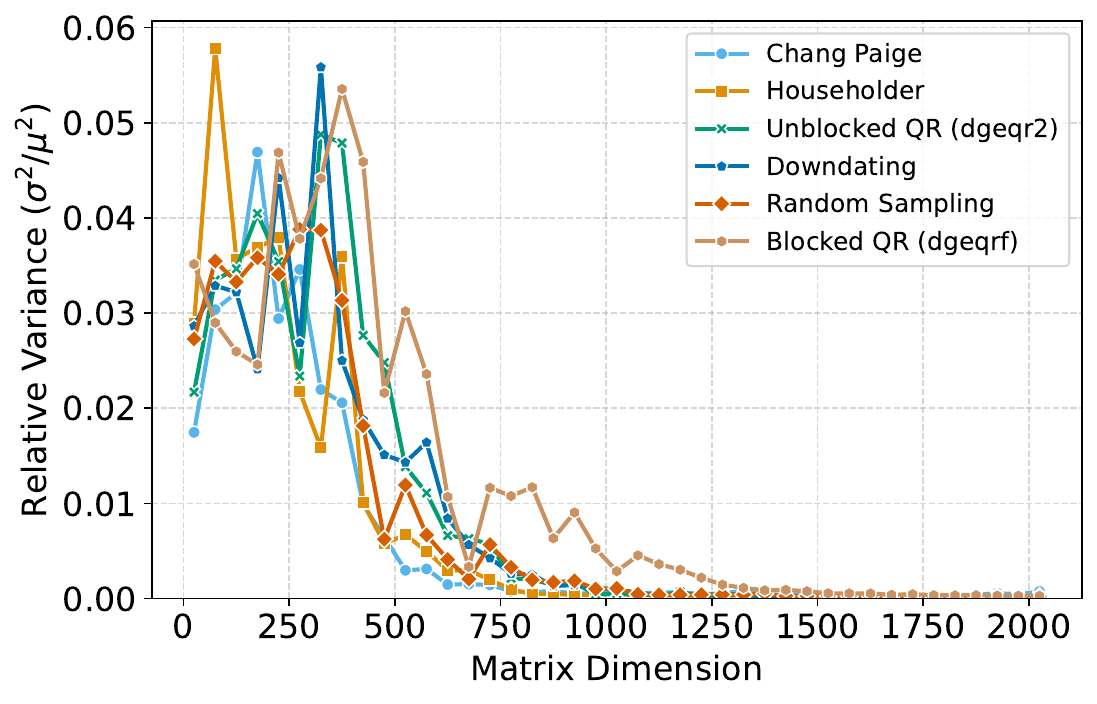}\caption{Variances for running times of exact and approximate V-BLAST algorithms,
as well as of two QR factorizations included for calibration from
figure~\ref{fig:perm-time-vs-size}. Each marker represents an average
of 100 runs on different Gaussian random matrices. The results are
normalized by the average running time to be unit less.}\label{fig:variance-vs-dim}
\end{figure}

\FloatBarrier

\paragraph*{Acknowledgments}

We thank Xiao-Wen Chang for enlightening discussions regarding the
V-BLAST algorithm and for sending us his implementation of the algorithm
that he and Chris Paige developed. AI was used in preparing this paper,
predominantly in formalizing the proofs, fixing issues in them, and
writing code for the experiments. The statements and proofs of the
claims in Section~\ref{sec:theoretical} were prepared by the authors,
with some back and forth with AI to improve and formalize the text.
The results were also verified in Lean using Claude. The entire manuscript,
up to small grammatical changes, was written by the authors. The authors
assume responsibility for all content.

\bibliographystyle{siamplain}
\bibliography{references}

@article{doi:10.1137/15M1044680,
author = {Duersch, Jed A. and Gu, Ming},
title = {Randomized {QR} with Column Pivoting},
journal = {SIAM Journal on Scientific Computing},
volume = {39},
number = {4},
pages = {C263-C291},
year = {2017},
doi = {10.1137/15M1044680},
}

@article{doi:10.1137/13092157X,
author = {Demmel, James W. and Grigori, Laura and Gu, Ming and Xiang, Hua},
title = {Communication Avoiding Rank Revealing QR Factorization with Column Pivoting},
journal = {SIAM Journal on Matrix Analysis and Applications},
volume = {36},
number = {1},
pages = {55-89},
year = {2015},
doi = {10.1137/13092157X},
}

@article{doi:10.1016/j.apnum.2007.01.008,
  title = {Euclidean distances and least squares problems for a given set of vectors},
  author = {Chang,  Xiao-Wen and Paige,  Christopher C.},
  volume = {57},
  doi = {10.1016/j.apnum.2007.01.008},
  number = {11-12},
  journal = {Applied Numerical Mathematics},
  year = {2007},
  pages = {1240–1244}
}

@article{doi:10.1179/1752270611y.0000000004,
  title = {Integer estimation methods for {GPS} ambiguity resolution: an applications oriented review and improvement},
  author = {Xu,  Peiliang and Shi,  Chuang and Liu,  Jingnan},
  volume = {44},
  doi = {10.1179/1752270611y.0000000004},
  number = {324},
  journal = {Survey Review},
  year = {2012},
  pages = {59–71}
}

@inproceedings{doi:10.1109/issse.1998.738086,
  title = {{V-BLAST}: an architecture for realizing very high data rates over the rich-scattering wireless channel},
  author = {Wolniansky,  P.W. and Foschini,  G.J. and Golden,  G.D. and Valenzuela,  R.A.},
  series = {ISSSE-98},
  doi = {10.1109/issse.1998.738086},
  booktitle = {1998 URSI International Symposium on Signals,  Systems,  and Electronics. Conference Proceedings (Cat. No.98EX167)},
  pages = {295-300},
  collection = {ISSSE-98},
  year = {1998},
}

@article{doi:10.1049/el:20010899,
  author = {W{\"u}bben,  D. and B{\"o}hnke,  R. and Rinas,  J. and K{\"u}hn,  V. and Kammeyer,  K.D.},
  title = {Efficient algorithm for decoding layered space-time codes},
  volume = {37},
  doi = {10.1049/el:20010899},
  number = {22},
  journal = {Electronics Letters},
  year = {2001},
  pages = {1348--1350}
}

@inproceedings{doi:10.1109/icassp.2000.859065,
  title = {An efficient square-root algorithm for BLAST},
  author = {Hassibi,  B.},
  series = {ICASSP-00},
  volume = {2},
  doi = {10.1109/icassp.2000.859065},
  booktitle = {2000 IEEE International Conference on Acoustics,  Speech,  and Signal Processing. Proceedings (Cat. No.00CH37100)},
  pages = {II737-II740},
  collection = {ICASSP-00},
  year = {2000},
}

@misc{doi:10.48550/arxiv.2505.12980,
  title = {Algorithms for Nonlinear Mixed-Integer Location Estimation},
  author = {Uziel,  Ophir and Fogel,  Efi and Halperin,  Dan and Toledo,  Sivan},
  doi = {10.48550/ARXIV.2505.12980},
  year = {2025},
}

@article{doi:10.1109/tit.2003.817444,
  title = {On maximum-likelihood detection and the search for the closest lattice point},
  author = {Damen,  M.O. and El Gamal,  H. and Caire,  G.},
  doi = {10.1109/tit.2003.817444},
  number = {10},
  journal = {IEEE Transactions on Information Theory},
  year = {2003},
  pages = {2389–2402}
}

@article{doi:10.1109/tcomm.2003.809789,
  title = {Achieving near-capacity on a multiple-antenna channel},
  author = {Hochwald,  B.M. and ten Brink,  S.},
  volume = {51},
  doi = {10.1109/tcomm.2003.809789},
  number = {3},
  journal = {IEEE Transactions on Communications},
  year = {2003},
  pages = {389–399}
}

@article{doi:10.1007/bf02579403,
  title = {On {Lovász}’ lattice reduction and the nearest lattice point problem},
  author = {Babai,  L.},
  volume = {6},
  doi = {10.1007/bf02579403},
  number = {1},
  journal = {Combinatorica},
  year = {1986},
  pages = {1–13}
}

@article{doi:10.1007/bf01581144,
  title = {Lattice basis reduction: Improved practical algorithms and solving subset sum problems},
  author = {Schnorr,  C. P. and Euchner,  M.},
  volume = {66},
  doi = {10.1007/bf01581144},
  number = {1-3},
  journal = {Mathematical Programming},
  year = {1994},
  pages = {181–199}
}

@article{doi:10.1007/bf01457454,
  title = {Factoring polynomials with rational coefficients},
  author = {Lenstra,  A. K. and Lenstra,  H. W. and Lovász,  L.},
  volume = {261},
  doi = {10.1007/bf01457454},
  number = {4},
  journal = {Mathematische Annalen},
  year = {1982},
  pages = {515–534}
}

@article{doi:10.1109/tit.2013.2253596,
  title = {Effects of the {LLL} Reduction on the Success Probability of the {Babai} Point and on the Complexity of Sphere Decoding},
  author = {Xiao-Wen Chang and Jinming Wen and Xiaohu Xie},
  volume = {59},
  doi = {10.1109/tit.2013.2253596},
  number = {8},
  journal = {IEEE Transactions on Information Theory},
  year = {2013},
  pages = {4915–4926}
}

@article{doi:10.1109/tit.2016.2627082,
  title = {Success Probability of the {Babai} Estimators for Box-Constrained Integer Linear Models},
  author = {Wen,  Jinming and Chang,  Xiao-Wen},
  volume = {63},
  doi = {10.1109/tit.2016.2627082},
  number = {1},
  journal = {IEEE Transactions on Information Theory},
  year = {2017},
  pages = {631–648}
}

@article{doi:10.1137/080731992,
  title = {Communication-optimal Parallel and Sequential {QR} and {LU} Factorizations},
  author = {Demmel,  James and Grigori,  Laura and Hoemmen,  Mark and Langou,  Julien},
  volume = {34},
  doi = {10.1137/080731992},
  number = {1},
  journal = {SIAM Journal on Scientific Computing},
  year = {2012},
  pages = {A206–A239}
}

@article{doi:10.1109/twc.2007.05806,
  title = {A Novel Fast Recursive {MMSE-SIC} Detection Algorithm for {V-BLAST} Systems},
  author = {Luo,  Zhendong and Liu,  Siyang and Zhao,  Ming and Liu,  Yuanan},
  volume = {6},
  doi = {10.1109/twc.2007.05806},
  number = {6},
  journal = {IEEE Transactions on Wireless Communications},
  year = {2007},
  pages = {2022–2025}
}

@article{doi:10.1109/tsp.2003.812897,
  title = {A fast recursive algorithm for optimum sequential signal detection in a {BLAST} system},
  author = {Benesty,  J. and Yiteng Huang and Jingdong Chen},
  volume = {51},
  doi = {10.1109/tsp.2003.812897},
  number = {7},
  journal = {IEEE Transactions on Signal Processing},
  year = {2003},
  pages = {1722–1730}
}

@article{doi:10.1109/twc.2009.081466,
  title = {On Fast Recursive Algorithms for {V-BLAST} with Optimal Ordered {SIC} Detection},
  author = {Shang,  Yue and Xia,  Xiang-gen},
  volume = {8},
  doi = {10.1109/twc.2009.081466},
  number = {6},
  journal = {IEEE Transactions on Wireless Communications},
  year = {2009},
  pages = {2860–2865}
}

@inproceedings{doi:10.1109/glocom.2005.1578079,
  title = {Low complexity adaptation of {MIMO} {MMSE} receivers,  implementation aspects},
  author = {Szczeci{\'n}ski,  L. and Massicotte,  D.},
  doi = {10.1109/glocom.2005.1578079},
  booktitle = {GLOBECOM ’05. IEEE Global Telecommunications Conference,  2005.},
  year = {2005},
}

@article{doi:10.1561/0400000060,
  title = {Sketching as a Tool for Numerical Linear Algebra},
  author = {Woodruff,  David P.},
  volume = {10},
  doi = {10.1561/0400000060},
  journal = {Foundations and Trends{\textregistered} in Theoretical Computer Science},
  year = {2014},
  pages = {1–157}
}

@article{doi:10.1006/jpdc.1996.0107,
  title = {{Cilk}: An Efficient Multithreaded Runtime System},
  author = {Blumofe,  Robert D. and Joerg,  Christopher F. and Kuszmaul,  Bradley C. and Leiserson,  Charles E. and Randall,  Keith H. and Zhou,  Yuli},
  volume = {37},
  doi = {10.1006/jpdc.1996.0107},
  journal = {Journal of Parallel and Distributed Computing},
  year = {1996},
  pages = {55–69}
}

@inproceedings{doi:10.1145/800076.802486,
  title = {{I/O} complexity: The red-blue pebble game},
  author = {Hong, Jia-Wei and Kung, H. T.},
  series = {STOC ’81},
  doi = {10.1145/800076.802486},
  booktitle = {Proceedings of the thirteenth annual ACM symposium on Theory of computing  - STOC ’81},
  year = {1981},
  pages = {326-333},
  collection = {STOC ’81},
}

@inbook{doi:10.1090/dimacs/050/09,
  title = {A survey of out-of-core algorithms in numerical linear algebra},
  author = {Toledo,  Sivan},
  doi = {10.1090/dimacs/050/09},
  booktitle = {External Memory Algorithms},
  year = {1999},
}

@article{doi:10.1137/090769156,
  title = {Minimizing Communication in Numerical Linear Algebra},
  author = {Ballard,  Grey and Demmel,  James and Holtz,  Olga and Schwartz,  Oded},
  volume = {32},
  doi = {10.1137/090769156},
  number = {3},
  journal = {SIAM Journal on Matrix Analysis and Applications},
  year = {2011},
  pages = {866–901}
}

@article{doi:10.1137/0910005,
  title = {A Storage-Efficient {$WY$} Representation for Products of {Householder} Transformations},
  author = {Schreiber,  Robert and Van Loan,  Charles},
  volume = {10},
  doi = {10.1137/0910005},
  number = {1},
  journal = {SIAM Journal on Scientific and Statistical Computing},
  year = {1989},
  pages = {53–57}
}

@book{doi:10.1137/1.9780898719604,
  title = {{LAPACK} Users’ Guide},
  author = {Anderson,  E. and Bai,  Z. and Bischof,  C. and Blackford,  L. S. and Demmel,  J. and Dongarra,  J. and Du Croz,  J. and Greenbaum,  A. and Hammarling,  S. and McKenney,  A. and Sorensen,  D.},
  doi = {10.1137/1.9780898719604},
  year = {1999},
}

@article{doi:10.1145/77626.79170,
  title = {A set of level 3 basic linear algebra subprograms},
  author = {Dongarra,  J. J. and Du Croz,  Jeremy and Hammarling,  Sven and Duff,  I. S.},
  volume = {16},
  doi = {10.1145/77626.79170},
  number = {1},
  journal = {ACM Transactions on Mathematical Software},
  year = {1990},
}

@article{doi:10.1145/42288.42291,
  title = {An extended set of {FORTRAN} basic linear algebra subprograms},
  author = {Dongarra,  Jack J. and Du Croz,  Jeremy and Hammarling,  Sven and Hanson,  Richard J.},
  volume = {14},
  doi = {10.1145/42288.42291},
  number = {1},
  journal = {ACM Transactions on Mathematical Software},
  year = {1988},
}

@book{doi:10.1137/1.9780898718027,
  title = {Accuracy and Stability of Numerical Algorithms: Second Edition},
  author = {Higham,  Nicholas J.},
  doi = {10.1137/1.9780898718027},
  year = {2002},
}

@book{doi:10.1137/1.9781611977523,
  title     = {Rounding Errors in Algebraic Processes},
  author    = {Wilkinson, James Hardy},
  year      = {1963},
  note      = {Republished by SIAM, 2023},
  doi       = {10.1137/1.9781611977523}
}

\end{document}